\documentclass[article,superscriptaddress,showpacs,aps,twocolumn,prb,floatfix]{revtex4}
\usepackage{eurosym}
\usepackage{graphicx,bm,amsmath,amssymb}
\usepackage[usenames]{color}

\newcommand{\bea}{\begin{eqnarray}}
\newcommand{\eea}{\end{eqnarray}}

\def\beq{\begin{equation}}
\def\eeq{\end{equation}}

\newcommand{\bR}{\mathbf{R}}

\newcommand{\rK}{\mathrm{K}}

\begin{document}

\title{Quantifying the leading role of the surface state in the Kondo effect of Co/Ag(111)}
\author{M. Moro-Lagares}
\affiliation{Instituto de Nanociencia de Arag\'on, Laboratorio de Microscopias Avanzadas,
	University of Zaraqoza, E-50018 Zaragoza, Spain}
\affiliation{Institute of Physics, Academy of Sciences, Prague, Czech Republic}
\affiliation{Regional Centre of Advanced Technologies and Materials, Faculty of Science,
	Department of Physical Chemistry, Palacky University, Olomouc, Czech Republic}
\author{J. Fern\'andez}
\affiliation{Centro At\'omico Bariloche and Instituto Balseiro, Comisi\'on Nacional deEnerg\'{\i}a At\'omica, 8400 Bariloche, Argentina}
\author{P. Roura-Bas}
\affiliation{Centro At\'omico Bariloche and Instituto Balseiro, Comisi\'on Nacional deEnerg\'{\i}a At\'omica, 8400 Bariloche, Argentina}
\author{M.R. Ibarra}
\affiliation{Instituto de Nanociencia de Arag\'on, Laboratorio de Microscopias Avanzadas,
	University of Zaraqoza, E-50018 Zaragoza, Spain}
\affiliation{Departamento F\'{\i}sica Materia Condensada,
	University of Zaraqoza, E-50018 Zaragoza, Spain}
\author{A. A. Aligia}
\email{aligia@cab.cnea.gov.ar}
\affiliation{Centro At\'omico Bariloche and Instituto Balseiro, Comisi\'on Nacional deEnerg\'{\i}a At\'omica, 8400 Bariloche, Argentina}
\author{D. Serrate}
\email{serrate@unizar.es}
\affiliation{Instituto de Nanociencia de Arag\'on, Laboratorio de Microscopias Avanzadas,
	University of Zaraqoza, E-50018 Zaragoza, Spain}
\affiliation{Departamento F\'{\i}sica Materia Condensada,
	University of Zaraqoza, E-50018 Zaragoza, Spain}
\date{\today }

\begin{abstract}
  Using a combination of scanning tunneling spectroscopy and atomic lateral manipulation, we obtained a systematic variation of the Kondo temperature ($T_\mathrm K$) of Co atoms on Ag(111) as a function of the surface state contribution to the total density of states at the atom adsorption site ($\rho_s$). By sampling the $T_\mathrm K$ of a Co atom on positions where $\rho_s$ was spatially resolved beforehand, we obtain a nearly linear relationship between both magnitudes. We interpret the data on the basis of an Anderson model including orbital and spin degrees of freedom (SU(4)) in good agreement with the experimental findings. The fact that the onset of the surface band is near the Fermi level is crucial to lead to the observed linear behavior. In the light of this model, the quantitative analysis of the experimental data evidences that at least a quarter of the coupling of Co impurities with extended states takes place through the hybridization to surface states. This result is of fundamental relevance in the understanding of Kondo screening of magnetic impurities on noble metal surfaces, where bulk and surface electronic states coexist.

\end{abstract}

\maketitle

\section{Introduction}\label{intro}

Single atoms with partially filled $d$- or $f$-shells on a solid state surface are known to exhibit strong electron correlations leading to a wide range of physical ground states. The magnetic properties of such impurities on metals are inherently connected with many-body interactions between the localized magnetic moment and the conduction electrons \cite{kondo64,madhavan01,knorr02,kouwenhoven01,henzl07,franke11, spinelli15,martinez17, cornils17}. In this framework, the Kondo effect \cite{kondo64,kondo68,hewson97} is the most frequently found. Since this phenomenon is an archetypal example of the formation of a many-body quantum state, it is central in the understanding of the electronic behavior of complex strongly correlated electrons systems such as heavy fermions \cite{hewson97,andres75}, Kondo insulators\cite{aeppli92}, and nanoscale systems \cite{roch08,parks10,florens11,vincent12,li98,madhavan98,manoharan00,madhavan01,knorr02,limot05,henzl07,serrate14,Zhao05,komeda11,minamitani12,iancu16,ormaza17}.

Thanks to the large spatial and energy resolution of scanning tunneling microscopy (STM) and spectroscopy (STS) \cite{madhavan98,madhavan01,li98}, these tools are extremely well suited to access the spectroscopic features of adsorbate induced many-body resonances in tunneling differential conductance ($dI/dV$). Most of STM studies on Kondo impurities are performed on noble metal (111) surfaces, where both bulk and surface electrons coexist \cite{madhavan98,manoharan00,madhavan01,knorr02,limot05,henzl07,serrate14,Zhao05,komeda11,minamitani12,iancu16,ormaza17,fernandez16}. Unavoidably, the question of whether surface or the bulk electrons play the leading role in the Kondo effect raises. To date, the answer remains unclear because there are conflicting conclusions depending on the technical approach to the problem. Since bulk electrons decay much faster than surface state electrons into the crystal, it has been common practice to measure the Kondo resonance as a function of the lateral distance to the atom \cite{ujsaghy00,knorr02,plihal01,henzl07}.

For instance,  Henzl {\itshape et al.} \cite{henzl07} concluded that bulk electrons determine the Kondo temperature ($T_\mathrm K$) of Co/Ag(111) by intentionally depleting the spectral weight of the surface state at Fermi level. The study of the Kondo resonance next to a monoatomic step edge led to the conclusion that the role of the surface states is marginal \cite{limot05}. This is supported by the weak dependence of $T_\mathrm K$ of Co on noble metal surfaces\cite{schneider05} with marked differences in the weight of their surface states relative to the bulk ones. On the contrary, the theoretically predicted \cite{ujsaghy00,plihal01} oscillations of the resonance line shape as a function of the tip lateral displacement on the order of the bulk electrons Fermi wavelength have not been observed\cite{knorr02,henzl07,madhavan01}. In fact, the theoretical description by Merino {\itshape et al.} \cite{merino04} cannot explain the distance dependent data on Co/Cu(111)\cite{knorr02} without a major involvement of the surface states.

The seminal work about the quantum mirage of the Kondo resonance into the focus of elliptical resonators proves unambiguously a finite contribution of surfaces states\cite{manoharan00}. Based on the relative intensity of $dI/dV$ at both foci (one with a Co impurity and
the other empty) a lower bound of 1/10 for the relative contribution of surface states has been estimated \cite{aligia05}. Moreover, the rather high $T_\mathrm K \sim 180$ K of a Co porphirine on ($\sqrt{3}\times\sqrt{3}$)Ag-Si(111), where bulk electrons states are not present, indicates that a significant coupling between the surface states and magnetic impurity is possible\cite{li09}. In support of this, it has been recently shown that $dI/dV$ of Ag(111) oscillates as the resonance width of Co atoms near step edges, quantum resonators or another atom\cite{li18}. It is worth noting that, from the theoretical point of view, the Kondo effect is extremely sensitive to the hybridization channels between the impurity and the metal host electrons, which exhibit non-trivial dependencies on the k-space electronic structure of the surface and the actual adsorption geometry \cite{lin05}. Thus, direct comparison of the Kondo resonance among different environments of the same adatom is physically inaccurate.

In this Article, we quantify the role of surface electron states in the Kondo effect of Co adatoms on Ag(111). We characterize their Kondo spectral features while varying just one single parameter of the problem: the surface state contribution to the local density of states of the substrate, $\rho_s$. In sections \ref{sym} and \ref{model} we develop the theoretical background on the basis of an Anderson model with SU(4) symmetry, which is consistent with the experimental spectroscopy as opposed to the SU(2) one\cite{li18}. Section \ref{results} is devoted to the experimental differential conductance $dI/dV$ at position $\bR$ with ($G_K$) and without ($G$) Co impurity between the tip and the Ag(111) surface. The analysis of $T_\mathrm K(\bR)$ and the amplitude of the Kondo resonance reveals that both magnitudes increase monotonically with $G(\bR)$. The theoretical calculation of the energy resolved $G$ for varying $\rho_s$ is given in section \ref{res}, using both the non-crossing approximation (NCA) and poor man's scaling (PMS). Finally, in section \ref{dicussion} the experimental and theoretical physical parameters are compared. We show that the coupling of the Co impurity state with extended states steaming from the surface state could be the dominant one, and prove a threshold of at least one fourth of that from the bulk states.

\section{Symmetry Analysis}\label{sym}

\begin{figure}[h!]
\includegraphics{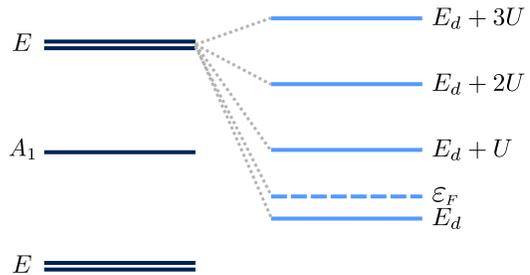}
\caption{Left: scheme of the splitting of the (one-particle) $3d$ orbitals
under the point group $C_{3v}$. Right: scheme of the splitting of the four antibonding states of
symmetry $E$ by the Coulomb repulsion. $\varepsilon_F$ denotes the position of
the Fermi energy compatible with the position of the observed Fano antiresonance.}
\label{scheme}
\end{figure}

In analogy with other nobel metal surfaces,\cite{ternes08} the Co atoms might occupy two non equivalent hollow positions on the Ag(111) surface, depending on whether the Co atoms lie above a Ag atom of the second layer or not (fcc/hcp). In both cases the symmetry point group is $C_{3v}$. This group has three irreducible representations: $A_{1}$ and $A_{2}$ of dimension one, and
the two dimensional representation $E$. Disregarding spin for the moment, the
Co $3d$ orbitals are split in one $A_{1}$ singlet and two $E$ doublets, as sketched at the left side of Fig. \ref{scheme}. Choosing the coordinates in such a way that $z$ is perpendicular to the
surface and one of the Ag atoms nearest to Co lies in the $xz$ plane, the $3d$
orbital with symmetry $3z^{2}-r^{2}$ transforms as the $A_{1}$
representation, $xz$ and $yz$ transform like the $E$ representation, and $x^{2}-y^{2}$ ($-xy$) transforms under the operations of $C_{3v}$ in the same
way as $xz$ ($yz$). Any Hamiltonian that respects the point group symmetry
(and without additional symmetry) mixes these two doublets, leading to
bonding and antibonding states. In particular, the antibonding $E$ states have
the form

\begin{eqnarray}
|e_1\rangle  &=&\alpha |xz\rangle +\beta |\left( x^{2}-y^{2}\right) /2\rangle ,
\notag \\
|e_2\rangle  &=&\alpha |yz\rangle -\beta |xy\rangle .  \label{e}
\end{eqnarray}%
Additional adatoms on the surface break the $C_{3v}$ symmetry, but this
effect is small if these atoms are sufficiently far from the Co atom under
study as is the case in this work.

The Coulomb repulsion inside the $3d$ orbitals splits the energy necessary to add
electrons in the same orbital. For example, let us call $E_d$ the energy necessary to add the
first electron in one of the antibonding $E$ orbitals with any spin. This energy does not depend
on the particular antibonding orbital chosen ($e_1$ or $e_2$) or its spin. However to add
the second electron, one has to pay the Coulomb repulsion $U$ between them. Similarly, the necessary energy to add the third or fourth electron is $E_d$ plus the Coulomb repulsion with the previous ones. This is presented schematically at the right of Fig. \ref{scheme}. The actual position of the levels is more complex because it is modified by exchange and pair hopping terms (see for example Ref. \citenum{aligia13}), but they not affect our treatment. For example, for the ground state for occupancy 2 in the antibonding $E$ is a triplet due to Hund's rules. Instead, for occupancy 3 of theses states the ground state is degenerate and is formed by two spin doublets with one hole in either $e_1$ or $e_2$. A similar splitting takes place for the bonding $E$ and the $A_1$ states, which remain occupied in the neutral Co atom.

While symmetry alone cannot determine the ordering of the levels, the position of the observed
Fano-Kondo dip $\omega _{\mathrm K}$ for positive energies of the order of the Kondo
temperature $T_\mathrm K$ or larger (see for instance Figs. \ref{fig:fig_2}b and \ref{dip}) \cite{note3} points to an SU(4) Kondo system with occupancy near 1, as we show below.
This in consistent with the configuration $3d^7$ expected for a neutral
Co atom, with four electron occupying the bonding $E$ orbitals, two in the $A_1$ orbital
and the remaining electron in one antibonding $E$ orbitals (Fig. \ref{scheme}). Other possibilities can be disregarded. For example if both $A_1$ states were the highest in energy putting two holes there and one in the antibonding $E$ orbitals, the model presented in Section \ref{model} still holds after an electron-hole transformation in the antibonding $E$ orbitals, in which case the Kondo dip would be to the left of the Fermi energ (i.e., same differential conductance as in Fig. \ref{dip} but with opposite sign for $\omega$). Assuming a $3d^8$ configuration, one has two possibilities to obtain a Kondo state: i) two holes in the antibonding $E$ states, but in this case the Kondo dip would be centered at the Fermi level \cite{barral17}, ii) one hole in an $E$ state and one hole in an $A_1$ state. This is the case of Fe phtalocyanine on Au(111) which shows a two-stage Kondo effect with two features of different width at the Fermi energy,\cite{minamitani12} completely different from our case. We have not discussed above combinations of holes in bonding and antibonding $E$ orbitals because they are unlikely for Co.

Therefore two channels are necessary to describe the system and one-channel models
[like the ordinary one-channel SU(2)  Anderson or Kondo model] are ruled out.
One has in principle a spin SU(2) times orbital SU(2) model.
However, for large $U$ (we take $U\rightarrow \infty $ but this is not an essential approximation \cite{note2}) the symmetry is SU(4) (larger than SU(2)$\times $SU(2)),
including orbital and spin degeneracies.

\section{Model and Formalism}\label{model}

\subsection{Hamiltonian}

The Hamiltonian can be written as

\begin{eqnarray}
H &=&\sum_{ki\sigma }\varepsilon _{k}^{s}s_{ki\sigma }^{\dagger }s_{ki\sigma
}+\sum_{k\sigma }\varepsilon _{k}^{b}b_{ki\sigma }^{\dagger }b_{ki\sigma
}+E_{d}\sum_{\sigma }d_{i\sigma }^{\dagger }d_{i\sigma }+  \notag \\
&&+U\sum_{i\sigma \neq j\sigma ^{\prime }}d_{i\sigma }^{\dagger }d_{i\sigma
}d_{j\sigma ^{\prime }}^{\dagger }d_{j\sigma ^{\prime }}+\sum_{k\sigma
}V_{k}^{s}[d_{i\sigma }^{\dagger }s_{ki\sigma }+\text{H.c.}]+  \notag \\
&&+\sum_{k\sigma }V_{k}^{b}[d_{i\sigma }^{\dagger }b_{ki\sigma }+\text{H.c.}%
].  \label{ham}
\end{eqnarray}
where $d_{i\sigma }^{\dagger }$ creates an electron in the antibonding
orbital $|e_i\rangle$ with spin $\sigma $, and $s_{ki\sigma
}^{\dagger }$ ($b_{ki\sigma }^{\dagger }$) are creation operators for an
electron in the $k^{th}$ surface (bulk) conduction eigenstate with symmetry $%
i$ and spin $\sigma $.

We assume constant densities of bulk states $\rho _{b}$ extending in a wide
range from $-D$ to $D$, and $\rho _{s}$ extending from $D_{s}$ to $D$ $(|D_s|<D)$. As we shall show, the fact that the surface band begins abruptly near the Fermi level at $D_s=-67$ meV \cite{li97} (neglected in alternative treatments \cite{li18}) plays an essential role in the interpretation of the results. We also assume constant hybridizations $V_b=V_{k}^{b}$ and $V_s=V_{k}^{s}$. We believe that these assumptions are not crucial as long as the dependence of these parameters on energy is smooth in a range of a few times $T_{\mathrm K}$
around the Fermi energy. We define the couplings of the impurity state to bulk and surface state electrons as $\Delta _{b}=\pi \rho _{b}|V_{k}^{b}|^{2}$ and $\Delta _{s}=\pi \rho_{s}|V_{k}^{s}|^{2}$ respectively. Our work allows to experimentally determine
the ratio of these two quantities. $E_{d}$ is the energy of the relevant impurity state.

We solve the model using two techniques: non-crossing approximation (NCA, section \ref{ncab})
\cite{hewson97,bickers87} and poor man's scaling (PMS, section \ref{pmsr}) \cite{hewson97,anderson70} on the effective Coqblin-Schrieffer model. These approaches are known to reproduce correctly the relevant energy scale $T_\rK$ and its dependence on the Anderson parameters. In contrast to Numerical Renormalization Group in which the logarithmic discretization of the conduction band \cite{zitko11rg,vaugier07} broadens finite-energy features  \cite{zitko11,vaugier07},
and leads to inaccurate Kondo temperatures when a step in the conduction band is near the Fermi level, NCA correctly describes these features. For instance, the intensity and the width of the charge-transfer peak of the spectral density (the one near $E_{d}$) was found \cite{aligia15,fernandez18} in agreement with other theoretical methods \cite{fernandez18,pruschke89,logan98} and experiment \cite{konemann06}. The NCA works satisfactorily in cases in which the density of conduction states is not smooth \cite{kroha98}, including in particular a step in the conduction band \cite{fernandez17}. Furthermore, it has a natural extension to non-equilibrium
conditions \cite{wingreen94} and it is specially suitable for describing
satellite peaks of the Kondo resonance, as those observed in Ce systems \cite{reinert01,ehm07}, or away from zero bias voltage in non-equilibrium transport
\cite{tosi15,dinapoli14,rourabas09transp,rourabas09dyn}. Due to shortcomings of the approximation for finite $U$ \cite{pruschke89, haule01, tosi11}, we restrict our calculations to $U \rightarrow \infty$ but this is not an essential approximation in our case \cite{note2}.

The PMS is a perturbative approach that integrates out progressively a small
portion of the conduction states lying at the bottom and at the top of the
conduction bands, renormalizing the Kondo exchange coupling $J_K$\cite{hewson97,anderson70}.

\subsection{The STM tunneling conductance}

The differential conductance $dI/dV$ is proportional to the spectral density
of the mixed state $h_{i\sigma }(\bR_{t})$ at the position of the STM tip $\bR_t$\cite{aligia05}.

\begin{eqnarray}
G_K(eV)&=&dI/dV \propto \sum_{i \sigma} \rho _{h i \sigma}(eV), \notag\\
\rho _{h i \sigma}(\omega)&=&\frac{1}{2\pi j}[\tilde{G}_{h i \sigma}(\omega -j \epsilon)
-\tilde{G}_{h i \sigma}(\omega +j \epsilon )],
\notag\\
h_{i \sigma } (\bR_{t}) &=&\frac{1}{N}[s_{i \sigma }(\bR_{t})+p_{b}b_{i \sigma }(\bR_{t})+p_{d}d_{i \sigma}(\bR_{t})].
\label{didv}
\end{eqnarray}
where $V$ is the sample bias potential of the STM, $e$ the electron elementary charge,
$\tilde{G}_{h i \sigma }(\omega )=\langle \langle h_{i \sigma };h_{i \sigma }^{\dagger}\rangle \rangle_{\omega }$ is
the Green's function of $h_{i \sigma } (\bR_{t})$, $j$ is the imaginary unit,
$\epsilon $ is a positive infinitesimal, $N$ is a
normalization factor, $p_{b}$ is the ratio of the
tunneling matrix element between the STM tip and the bulk states $b_{i \sigma }$ and between
tip and surface states $s_{i \sigma }$, while $p_{d}$ is the analogous ratio for Co state
$d_{i \sigma }(\bR_{t})$ and surface states at the tip position.
$h_{i \sigma }(\bR_{t})$ represents the linear combination of surface, bulk and Co 3d states probed by the tip.

Using equations of motion, $\rho _{h i \sigma}(\omega )$ can be related with  the
Green's function for the $d$ electrons $\tilde{G}_{di\sigma }(\omega )=\langle
\langle d_{i \sigma };d_{i\sigma }^{\dagger }\rangle \rangle _{\omega }$, and
the unperturbed Green's functions for conduction/bulk electrons $\tilde{G}_{s/b}^{0}(\omega)$. In absence of magnetic and symmetry-breaking fields we can drop the subscripts $i\sigma$:
\begin{equation}
\tilde{G}_{h}(\omega ) =\tilde{G}_{s}^{0}(\omega )+p_{b}^{2}\tilde{G}_{b}^{0}(\omega )+\Delta
\tilde{G}_{h}(\omega )  \label{gh}
\end{equation}

$\Delta \tilde{G}_{h}(\omega)=0$ if the Co impurity is absent and if not

\begin{eqnarray}
\Delta \tilde{G}_{h}(\omega ) &=&F^{2}(\omega )\tilde{G}_{d}(\omega ),  \notag \\
F(\omega ) &=&V_{s}\tilde{G}_{s}^{0}(\omega )+p_{b}V_{b}\tilde{G}_{b}^{0}(\omega )+p_{d},
\label{deltagh}
\end{eqnarray}%
where

\begin{eqnarray}
\text{ }\tilde{G}_{b}^{0}(R_{i},R_{i},\omega ) &=&\rho _{b}\left[ \ln \left( \frac{%
\omega +D}{\omega -D,}\right) \right] , \notag \\
\tilde{G}_{s}^{0}(R_{i},R_{i},\omega ) &=&\rho _{s}\left[ \ln \left( \frac{\omega
-D_{s}}{\omega -D}\right) \right].  \label{g0}
\end{eqnarray}

\section{Experimental Results}\label{results}

Single Co atoms were deposited at low temperatures onto the Ag(111) surface ($T_{ev}\sim3$ K for an experimental temperature $T=1.1$ K) cleaned by repeated cycles of sputtering with Ar$^+$ and annealing at 500 $^\circ$C in UHV ($P_{base}\leq 1\times 10^{-10}$ mbar). We use a lock in amplifier to perform STS as a function of the applied sample bias, $V$. STS was acquired at constant height defined by the regulation set point $V_0$, $I_0$ on Ag(111) with rms modulation voltage $V_{mod}$ and implemented in two modes: (i) Single point $dI/dV$ spectroscopy ($V_{mod}=0.5$ mV, $V_0=-100$ mV, $I_0=42$ pA) on top of Co atoms to obtain the energy resolved $G_K(\bR)$; and (ii) $dI/dV(x,y)$ mapping at Fermi level ($V_{mod}=1$ mV, $V_0=-100$ mV, $I_0=200$ pA) to measure the spatially resolved $G(\bR)$ of the Ag(111) inspected area after clearing it away from atoms by means of atomic manipulation (typical set point for manipulation $V_0=3$ mV, $I_0=40-70$ nA). The working temperature is $T=1.1$ K or $T=4.7$ K, being the Kondo features of one atom in STS identical at both temperatures.

\begin{figure}[t]
	\centering
	\includegraphics{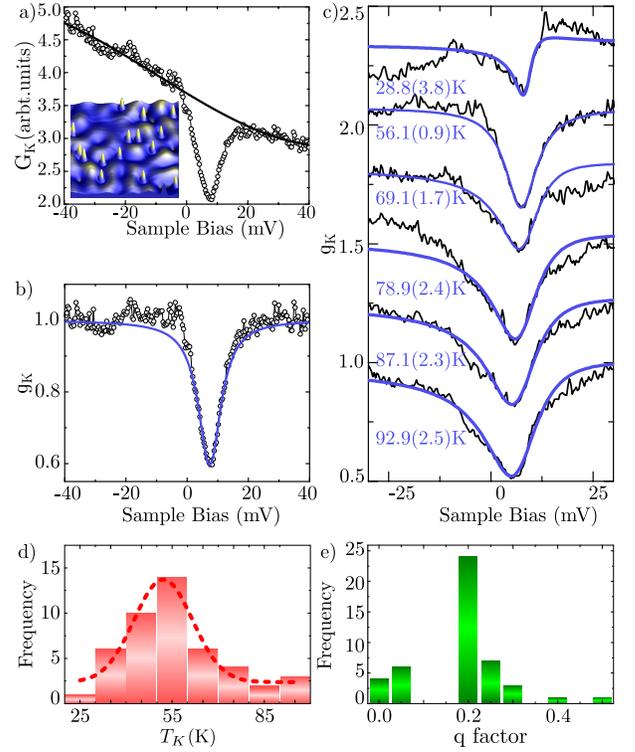}
	\caption{(a) Representative raw d$I$/d$V$ ($G_K(V)$, empty circles) showing the Kondo zero bias feature at the center of a single Co atom and background estimation (${\cal G}_0(V)$, solid line). (b) Fit of the resulting $g_K(V)$ to equation \ref{eq:rhok} yielding $T_\mathrm K=56.1\pm0.9$ K, $q=0$ and $\omega_K=7.39\pm0.04$ meV. (c) Dispersion found in the Kondo resonance of set of atoms spread on Ag(111) with the corresponding fit and $T_\mathrm K$ value. (d-e) Kondo Temperature and $q$ factor statistics. Using a Gaussian distribution profile (dashed line) for the $T_\mathrm K$ histogram, we obtain $\langle T_K\rangle=52.1\pm 9.4$ K.}
	\label{fig:fig_2}
\end{figure}

Experimentally, the Kondo effect of isolated Co atoms on metals manifests as a Fano resonance \cite{madhavan98,madhavan01,schneider02,schneider05} in the impurity $G_K$ near the Fermi level. We describe this singularity as $G_K={\cal G}_0g_K$, where ${\cal G}_0(\bR,\omega)$ is the convolution of the tip and the impurity density of states in absence of Kondo screening and $g_K(\bR,\omega)$ contains the Fano function, ${\cal F}(x,q)=(x+q)^2/1+x^2$, as follows:

\begin{equation}\label{eq:rhok}
g_K(\bR,\omega)=(1-A_K(\bR))+A_K(\bR){\cal F}\left[\frac{\omega-\omega_K}{\Gamma_0(\bR)/2},q\right]
\end{equation}

Here $\omega=eV$, $\omega_K$ the energy of the center of the Kondo resonance, $q$ the Fano asymmetry factor, $A_K(\bR)$ the resonance amplitude when the atom sits at surface position $\bR$, and $\Gamma_0(\bR)$ the resonance width,  which is related to the Kondo temperature as $2k_BT_\rK\simeq\Gamma_0$ for $T/T_\rK\rightarrow0$\cite{madhavan98,nagaoka02}. Below $T_\rK$ the spin of the extended states couples antiferromagnetically and screens the impurity spin, giving rise to the Kondo state\cite{hewson97,kondo68}. Figures \ref{fig:fig_2}(a-b) shows the analysis of a Kondo resonance based on Eq. \eqref{eq:rhok}, which permits to extract the parameters $T_\mathrm K(\bR)$, $A_K(\bR)$, $q$ and $\omega_K$ for each individual atom at position $\bR$.

 We first analyze $G_K$ of several Co atoms dispersed over the surface at their position right after the evaporation process (i.e., prior to any atom repositioning with the tip). Figures \ref{fig:fig_2}(c-e) unveil a significant uncertainty in the parameters describing the Kondo resonance. The histograms elaborated from a set of 40 different atoms are shown in Figures \ref{fig:fig_2}(d-e). $T_K$ spans over a range of 28 K $\leq T_\rK \leq 95$ K, with $\langle T_\mathrm K\rangle=52.1\pm9.4$ K being the most probable value. The most frequently found value for $A_K$ and $q$ is 0.2.

Apart from the \emph{hcp/fcc} character of the hollow sites in a (111) surface termination, the adsorption geometry of disperse Co atoms is indistinguishable. We have confirmed that the Kondo parameters are the same in both sites except for a slightly lower amplitude $A_K$ in one of them. Therefore, the different values obtained for
$T_K$ and $q$ suggest a sensitivity to the density of surface states
$\rho_s({\mathbf R})$. Particularly, in Ag(111), the onset of surface state lies in close proximity ($D_s=-67$ meV) to the Fermi level\cite{li97}, leading to a Fermi wavelength $\lambda_F\sim8$ nm\cite{fernandez16}, which is comparable to the distance between surface scatterers such as step edges, point impurities or Co adatoms. This will produce interference patterns in $\rho_s(\bR)$ with a characteristic length scale of $\lambda_F/2$. We have shown elsewhere\cite{fernandez16} that $\rho_s$ contributes strongly to the total density of states $h_{i \sigma }(\bR_{t})$ ($\rho_h$)
probed by the tip. Therefore, it is natural to expect that changes in $\rho_s(\bR)$ lead to the observed dispersion of $T_K$ of Co/Ag(111), through the hybridization of the Co 3d electrons with the surface states. This will become clear in Section \ref{pmsr}, where an analytical expression for the dependence of $T_K$ with $\rho_s$ is presented.

 \begin{figure}[]
 	\centering
 	\includegraphics[width=1\linewidth]{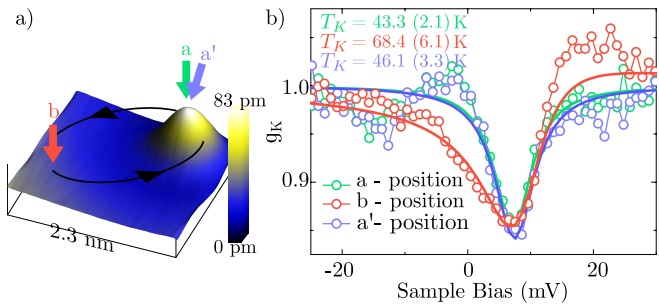}
 	\caption{a) Representation of the experiment carried out in the study of the variations of the Kondo resonance with the  point contact $\rho_s$.  b) d$I$/d$V$ spectrum (circles) and fit to Eq. \eqref{eq:rhok} of the Co atom at $a-$position (green); $b-$position (red) and $a'-$position (blue). $a'-$position is the same as the $a-$position but after atomic manipulation.}
 	\label{fig:fig_3}
 \end{figure}

To benchmark the correlation of $T_K$ with the electronic properties of the substrate, we measure $g_K$ (see Eq. \eqref{eq:rhok}) over a Co atom at its natural adsorption site $\bR$, and subsequently at another position $\bR^\prime$ far enough as to have presumably a different $\rho_s$ ($|\bR-\bR^\prime|\sim\lambda_F/2$). In Fig. \ref{fig:fig_3} we show $g_K$ at each site and in the absence of any tip change during the manipulation procedure. We find a strong variation of $\Delta T_K=T_\mathrm K(\bR^\prime)-T_\mathrm K(\bR)=24\pm4$ K, well above the experimental uncertainty. This experiment shows unambiguously that the coupling strength between the localized spin and the Fermi gas of conduction electrons is strongly influenced by the local value of $\rho_s$  at each contact point.

\begin{figure*}[]
	\centering
	\includegraphics{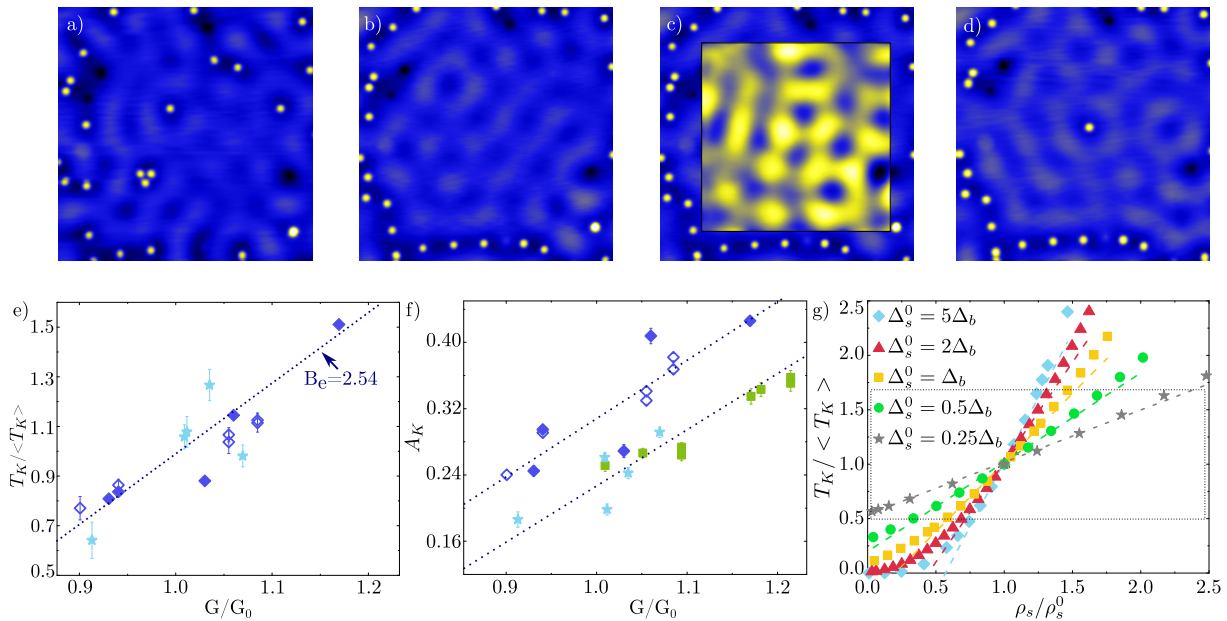}
	\caption{$T_K$ and $A_K$ variations with $G({\mathbf R})$. a) STM image of Co/Ag(111) after Co deposition.
	b) Co atoms are removed from the working area to avoid their influence on $G$.
	c) Inset: constant height $G({\mathbf R})=dI/dV$ map taken at $V=3$ mV ($I_0=200$ pA, $V_0=-100$ mV, and $V_{mod}=1$ mV).
	d) Co atom relocated at a certain position over the
	surface. e) Experimental dependence of $T_K/\langle T_K \rangle$ on the normalized local tunneling conductance, $G/G_0$. f) Experimental dependence of $A_K$ on $G/G_0$. In e) and f), the color-code represents data sets taken with the same tip, while for the same color, the opened/closed circle-code distinguishes data sets at two nearby different working areas. All measurements were taken in constant height mode at $T$=1.1 K ($I_0=42$ pA, $V_0=-100$ mV, and $V_{mod}=0.5$ mV). g) Theoretical dependence of $T_K/\langle T_K \rangle$ on $\rho_s/\rho^0_s$ for different values of $\Delta_s^0/\Delta_b$. The dotted lines are linear fits in the region $0.5 < T_K /\langle T_K \rangle <1.5$. The region enclosed within the dotted rectangle corresponds to the experimental parameter range.}	\label{fig:fig_4}
\end{figure*}

Next, we evaluate more precisely this position dependent Kondo effect through the analysis of $T_\rK$ and $A_K$  of Co atoms relocated in a region where $G(\bR)$ at Fermi level has been previously characterized (without Co atoms) in constant height conditions. First, we clean out the atoms in the selected working area as depicted in Figs. \ref{fig:fig_4}(a,b). Second we take a $G(\bR)$ image of the differential conductance near Fermi level ($0<V<3$ mV) as shown in Fig. \ref{fig:fig_4}c, whose maxima and minima reflect the characteristic interference pattern of the surface state. Afterwards, a single Co atom is moved across the inspected Ag(111) area (Fig. \ref{fig:fig_4}d) and we measure its energy spectrum $G_K(\bR)$ for each $\bR$ location. Note that this procedure is free of feedback artifacts, and that the drift between consecutive images is corrected by referring always $\bR$ to a reference feature of the same image.

At the tip-sample distance at which the experiment is performed the STM does not exhibit atomic resolution. Thus, $G(\bR)$ oscillations are only contributed by $\rho_s(\bR)$, owing to the interference pattern of scattered surface state quasiparticles. In Figs. \ref{fig:fig_4}(e,f) we plot $T_K/\langle T_K \rangle$ and $A_K$ as a function of $G/G_0$ for four different data sets gathered together, taken with different tips (symbol code) at different working areas (color code). $G_0$ is defined as the tunneling conductance of an ideal surface without scattering sources. Experimentally, we determine $G_0$ as the average differential conductance at Fermi level of a region much larger than $\lambda_F$, as the one shown in Fig. $\ref{fig:fig_4}c$. This normalization makes the analysis insensitive to the specific electronic structure of the tips used for the experiment. The resulting graphs display a monotonic increase of $T_\mathrm K$ and $A_K$ with $G$, which implicitly provides an evidence of the linear dependence of these parameters on $\rho_s$ within the experimental boundaries.

\section{Theoretical Results}\label{res}

In this section we present the theoretical results for the dependence of $T_\rK$ on the surface states density, $\rho_s$. For
simplicity, from now on we choose the origin of energies at $\varepsilon _{F}=0$. We have taken $D_{s}=-67$ meV from experiment \cite{limot05,phd:moro,fernandez16} and have
chosen $D=4$ eV, $\rho _{b}=0.135$ eV$^{-1}$, $\rho _{s}=0.0446$ eV$^{-1}$ (Ref. \citenum{fernandez16}). The results are rather insensitive to these parameters if the hybridizations are changed to  fix the  values of $\Delta _{b}$ and $\Delta _{s}$. For the energy of the occupied antibonding $E$ state with majority spin (see Fig. \ref{scheme}), we take
$|E_{d}|\gg\Delta_{s,b}$ (in particular $E_{d}=-2.2$). A different value would  simply require a rescaling of $\Delta_{s,b}$.

Concerning the parameters entering Eq. (\ref{didv}), previous comparison
between experiment and theory on the action of Co resonators on the surface
states\cite{fernandez16} suggest that $p_{b}^{2}\approx 1/16$.
At first we have taken $|p_{b}|=1/4$, but this implies a very large surface contribution
($\Delta_s^0/\Delta_b > 6$, see below). Furthermore, this estimation applies to a different tunneling barrier height\cite{fernandez16}, which may strongly alter the ratio $p_b$.
Therefore we think that it is better to be cautious and treat $p_b$ as an unknown parameter.
The shape of the resulting differential conductance $dI/dV$ is
rather insensitive to the sign of $p_{b}$ but the intensity is smaller for $p_{b}<0$.
The parameter $p_{d}$ is determined by fitting the line shape.
The line shape is rather insensitive to $p_{b}$ if $p_{d}$ is adjusted.

\subsection{Non-crossing approximation}

\label{ncab}

\subsubsection{Calculation of the Kondo temperature}

To determine theoretically the value of the Kondo temperature $T_\rK$, we calculate the conductance through the magnetic impurity as a function of temperature $G_d(T)$
for a hypothetical case with $p_d \rightarrow \infty$
and look for the temperature such that $G_d(T_{K})=\gamma_{0}/2$, where $\gamma_{0}$ is the ideal
conductance of the system (reached for $T=0$ and occupancy 1 of the impurity
level). Alternative definitions of $T_\rK$ differ in factor of the order of
1\cite{tosi12}, which is not relevant to us, as we shall show. We are interested in the
dependence of $T_\rK$ with $\Delta _{s}$. In practice we take

\begin{equation}\label{conductance}
 G_d(T) =  \gamma_0 \frac{\pi\Delta}{2}\int ~d\omega (-\frac{\partial f(\omega)}{\partial \omega}) \rho_d(\omega),
\end{equation}
where $\Delta=\Delta_b+\Delta_s$, $\rho_d(\omega)= \sum_{i \sigma} \rho_{d i \sigma}(\omega)$
is the total impurity spectral density adding both orbitals $i$
and spins $\sigma$, and $f(\omega)$ is the Fermi function.

\subsubsection{Fit of the experimental data}

\begin{figure}[h!]
\includegraphics{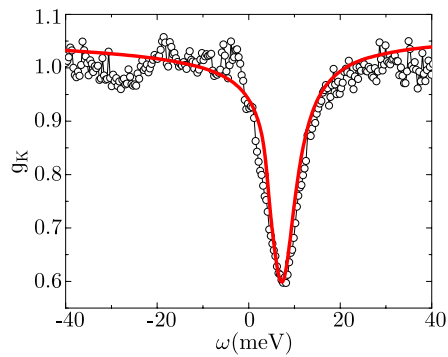}
\caption{(Color online) Differential conductance as a function of voltage.
Open circle: experimental $g_k$ (same as Fig. \ref{fig:fig_2}b) without background.
Red line: theory for $\Delta_s=69.26$ meV, $\Delta_b=256.5$ meV, $p_{b} = 1.16$, and  $p_{d}=7$.}
\label{dip}
\end{figure}

In  Fig. \ref{dip} we show one experimental result for the differential
conductance for which  the resulting $T_\rK$ is very near to the average one
$\langle T_\rK \rangle$, and the corresponding theoretical fit obtained at the experimental temperature $T=4.7$ K.
For the latter, we have assumed $\Delta_{s}= 0.27 \Delta _{b}$, $p_{b} = 1.16$, which is consistent with the experimental slope of $T_\rK$ vs. the tunneling conductance (see below) and adjusted $p_d$ to fit the experimental data. Very similar fits are obtained for larger values of $p_{b}$. The fit requires to shift the theoretical results by 4 meV to reach the experimental position of the dip $\omega_K=7.7$ meV. The reason of this discrepancy might be due to details on the energy
dependence of $\Delta _{s}$, which are particularly sensitive to the position
of the adatoms \cite{fernandez16} and we have neglected in our approach.

Note that for the parameters in Fig. \ref{dip}, the total width of the Fano dip is $\Gamma_0=8.71$ meV,
while twice $T_K$ obtained from the definition based on Eq. \eqref{conductance} gives $2k_\mathrm BT_\rK=9.78$ meV.
This ratio is approximately constant for the different parameters used here.
Our Fano fit for this experimental curve gives $T_K=56.1$ K $\sim$ 4.83 meV. Therefore we assume that
this value is representative of the average Kondo temperature $\langle T_K \rangle$
observed in experiment.
Note that the ratio $T_K/ \langle T_K \rangle$ does not depend on the
definition of $T_K$. We define  $\rho_{s}^0$ and $\Delta _{s}^0=\pi \rho_{s}^0 V^2_s$ as the values of the
surface spectral density and $\Delta _{s}$ that lead to $T_\rK=\langle T_\rK \rangle$.
$T_\rK$ depends mainly on $\Delta _{s} + \Delta _{b}$ and several ratios
$\Delta _{s}/\Delta _{b}$ can lead to the same $T_\rK$.

In Fig. \ref{fig:fig_4}g we show the dependence of $T_\rK/ \langle T_\rK \rangle$
vs. $\Delta_{s}/\Delta_{s}^0=\rho_{s}/\rho_{s}^0$ for several values of $R=\Delta_s^0/\Delta_b$. In good agreement with the experimental behavior of $T_\rK$ (Fig. \ref{fig:fig_4}e), we obtain a linear trend in the interval $0.5 < T_\rK/ \langle T_\rK \rangle < 1.5$ with slope $B$. As expected, $B$ increases with increasing $R=\Delta_s^0/\Delta_b$. For larger $R$ the linear dependence weakens and some curvature appears. The results for the slope $B$ for different ratios $R=\Delta_s^0/\Delta_b$ and what it implies for $p_{b}$ are listed in Table \ref{tab1}.

\begin{table}[ht!]
\caption{Slope of $T_K/ \langle T_K \rangle$ vs. $\rho_{s}/\rho_{s}^0$, the corresponding
$C_b$ value (see section \ref{dicussion}) and coefficient of the bulk density of states in Eq. (\ref{deltagh})
for different ratios $R=\Delta_s^0/\Delta_b$.}
\label{tab1}
\begin{tabular}{@{}cccc}
\hline
$R$   & $B$ & $C_b$ &  $p_{b}$\\
\hline\hline
0.25  & 0.480       	   & 4.289           		  & 1.190                              \\
0.27  & 0.503        	   & 4.045           		  & 1.156                               \\
0.5  & 0.820       	   & 2.098            		  & 0.832                              \\
1  & 1.269       	   & 1.001           		  & 0.575                              \\
2     & 1.835       	   & 0.384                        & 0.356          	               \\
5     & 2.375       	   & 0.070            		  & 0.152       	               \\
\hline
\end{tabular}
\end{table}

\subsection{Poor man's scaling}

\label{pmsr}

The PMS \cite{hewson97,anderson70} for this SU(4)  problem (or in general for SU($N$)
symmetry) up to second order in the Coqblin-Schrieffer interaction $J_K$ has
the same form as for the SU(2) Kondo Hamiltonian treated previously\cite{fernandez17},
taking $NJ_K$ as the interaction constant. Then, borrowing previous
results and taking the limit $U \rightarrow \infty$ we obtain the following analytical
formula for the Kondo temperature as a function of  $\Delta _{b}$ and $\Delta _{s}$

\begin{eqnarray}
T_{K} &\simeq &A|D_{s}|^{\eta }D^{1-\eta }\exp \left[ \frac{\pi
E_{d}}{4(\Delta _{b}+\Delta _{s})}\right] ,  \notag \\
\eta  &=&\frac{\Delta _{s}}{(\Delta _{b}+\Delta _{s})}.
\label{tkpms}
\end{eqnarray}

where for second order in $J_K$, $A=1$. Higher order corrections reduce $A$ and
introduce logarithmic corrections. However, in our case, it is not possible to obtain
an analytical formula like Eq. (\ref{tkpms}) if these corrections are included.

\begin{figure}[h!]
\includegraphics{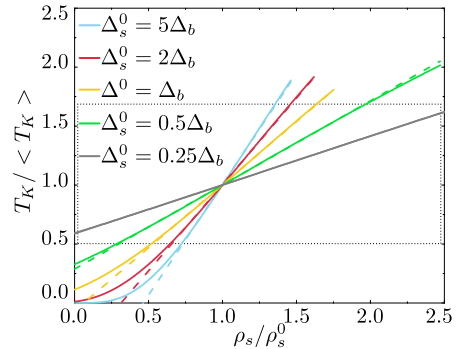}
\caption{(Color online) Same as Fig. \ref{fig:fig_4}g using Eq. (\ref{tkpms}). $T_{K}^{0}$ is the value obtained for the same parameters as before, and differs from the value of $\langle T_\rK \rangle$ by a factor $\sim 2$.  The region enclosed within the dotted rectangle corresponds to the experimental parameter range.}
\label{pms}
\end{figure}

In Fig. \ref{pms} we plot this function for the same parameters of Fig. \ref{fig:fig_4}g, showing again a linear dependence in the relevant range of parameters, in agreement with experiment. We obtain a semiquantitative agreement with the NCA (which assumes $U\rightarrow \infty $).

Eq. (\ref{tkpms}) sheds light  on the expected dependence of $T_\rK$ as a function of
$\Delta _{s}$.
In the experimentally relevant range of parameters the last (exponential)
factor has a marked upward curvature
which is largely compensated by the factor $|D_{s}|^{\eta }D^{1-\eta }$
leading to the approximately linear dependence displayed in Fig. \ref{pms}.
Replacing $|D_{s}|^{\eta }D^{1-\eta }$ by $D \sim 4$ eV (as in Ref. \onlinecite{li18})
cannot reproduce our experimental results.

In Table \ref{tab2} we display the slope ($B$) obtained from a linear fit in the interval $0.5 < T_K/ \langle T_K \rangle < 1.5$. The slope with NCA is about 13\% (for lower $R$) to 20\% (for larger $R$) larger than with PMS (cf. Tables \ref{tab1} and \ref{tab2}). The agreement might be improved including numerically logarithmic corrections of order $J_K^3$ but we failed in the attempt to calculate them.

\begin{table}[ht!]
\caption{Same as Table I calculated with PMS.}
\label{tab2}
\begin{tabular}{@{}cccc}
\hline
$R$   & $B$ & $C_b$ &  $p_{b}$\\
\hline\hline
0.25  & 0.414       	   & 5.135            		  & 1.302                              \\
0.27  & 0.445        	   & 4.709           		  & 1.247                              \\
0.5  & 0.713       	   & 2.562            		  & 0.920                              \\
1  & 1.070       	   & 1.374            		  & 0.674                              \\
2     & 1.465       	   & 0.734                       & 0.492          	               \\
5     & 1.878       	   & 0.352            		  & 0.341        	               \\
\hline
\end{tabular}
\end{table}

\section{Quantitative Discussion}\label{dicussion}

We have obtained experimentally and theoretically a linear trend of $T_\rK(\rho_s)$. This might be surprising at first sight, since an exponential dependence of $T_\rK$ with the coupling strength is expected \cite{hewson93,yang08}. However, due to the always existing bulk contribution, the proximity of the bottom of the surface band to the Fermi level and the particular region of interest of the parameter phase space (see the analytical PMS result in Eq. \eqref{tkpms}), the expected upward curvature is strongly reduced, particularly for small $R$.

Since the measurements are performed at constant height, the experimental $dI/dV$ without a Co impurity can be written for all positions as $G=C(\rho_{s}+p_{b}^{2} \rho_{b})$ (from Eqs. \eqref{didv} and \eqref{gh} with $\Delta \tilde{G}_{h}(\omega)=0$). Here
$C$ and  $\rho_{b}$ are constants. We write it in the form $G=C\rho_{s}^0(\rho_{s}/\rho_{s}^0+C_b)$, where $C_b=p_{b}^{2} \rho_{b}/\rho_{s}^0$ is the relative weight of the bulk states in tunneling conductance at reference point $T_K/ \langle T_K \rangle=1$. $C_b$  is also a constant. Now, the theoretical analogue of $G_0$ yields $G_0=C\rho_s^0(1+C_b)$.

To compare the values of our theoretical slope $B$ of $T_K/ \langle T_K \rangle$ vs. $\rho_{s}/\rho_{s}^0$, with the experimental slope $B_e \simeq 2.54$ of $T_K/ \langle T_K \rangle$ vs. $G/G_0$ obtained from the data in Fig. \ref{fig:fig_4}e, we must take into account that

\begin{eqnarray}
\frac{G}{G_0} = \frac{\rho_{s}/\rho_{s}^0+C_b}{1+C_b}.
\label{bb}
\end{eqnarray}

It can be readily shown that $B_e=(1+C_b)B$. The fact that $\rho_{s} \geq 0$ for the minimum $G$ observed, $G_{\mathrm min}/G_0=0.8$ (see Fig. \ref{fig:fig_4}), implies that $C_b/(1+C_b)<0.8$, which leads to the upper bound $\sim 4$ for $C_b$. The corresponding theoretical value of $B=0.503$ for the NCA method is obtained for $R=0.27$ (Table \ref{tab1}). Previously, a lower bound 0.1 was estimated for Co on Cu(111) based on the quantum mirage effect assuming $C_b=1$ \cite{aligia05}. For a more realistic value of the minimum $\rho_s$ about 60 \% of $\rho_s^0$ (the value for a surface without scattering sources), we obtain $B=1.269$ and $R=1$ (Table \ref{tab1}), i.e., the same coupling of the impurity to the surface states as to the bulk ones.

\section{Conclusions}

 By combining STS, atomic lateral manipulation, and applying a suitable Anderson Hamiltonian for the system, we have demonstrated that surface states have a relevant contribution in the formation of the Kondo state of Co/Ag(111). This result can be extended to other noble metal surfaces and provides an important clue in the understanding of more complex correlated electron systems. The sensitivity of $T_K$ to the surface state suggests the possibility to tune the coupling strength between a magnetic impurity and its foremost environment using confining nanostructures with size comparable to $\lambda_F$ \cite{fernandez16}. In the case of Co/Ag(111) we provide a lower bound for the coupling of surface states to Co 3d-states that is 27 \% of the one to bulk states. Furthermore, we show that a two-channel SU(4) Anderson model (considering both spin and orbital quantum numbers) is more appropriate to describe the Kondo effect than the one-channel SU(2) model. We also show that the proximity of the the surface density of states onset to the Fermi level plays a crucial role in the observed approximately linear dependence of the Kondo temperature with the surface density of states.

\section*{Acknowledgements}
We thank N. Lorente and R. Robles for fruitful discussions. We acknowledge financial support provided by the Spanish MINECO (grants MAT2013-46593-C6-3-P and MAT2016-78293-C6-6-R), as well as the Argentinian CONICET (PIP 112-201101-00832) and ANPCyT (PICT
2013-1045). M.M.L., D.S. and M.R.I. acknowledge the use of SAI-Universidad de Zaragoza.


\begin{thebibliography}{70}
\expandafter\ifx\csname natexlab\endcsname\relax\def\natexlab#1{#1}\fi
\expandafter\ifx\csname bibnamefont\endcsname\relax
  \def\bibnamefont#1{#1}\fi
\expandafter\ifx\csname bibfnamefont\endcsname\relax
  \def\bibfnamefont#1{#1}\fi
\expandafter\ifx\csname citenamefont\endcsname\relax
  \def\citenamefont#1{#1}\fi
\expandafter\ifx\csname url\endcsname\relax
  \def\url#1{\texttt{#1}}\fi
\expandafter\ifx\csname urlprefix\endcsname\relax\def\urlprefix{URL }\fi
\providecommand{\bibinfo}[2]{#2}
\providecommand{\eprint}[2][]{\url{#2}}

\bibitem[{\citenamefont{Kondo}(1964)}]{kondo64}
\bibinfo{author}{\bibfnamefont{J.}~\bibnamefont{Kondo}},
  \bibinfo{journal}{Progress of Theoretical Physics}
  \textbf{\bibinfo{volume}{32}}, \bibinfo{pages}{37} (\bibinfo{year}{1964}).

\bibitem[{\citenamefont{Madhavan et~al.}(2001)\citenamefont{Madhavan, Chen,
  Jamneala, Crommie, and Wingreen}}]{madhavan01}
\bibinfo{author}{\bibfnamefont{V.}~\bibnamefont{Madhavan}},
  \bibinfo{author}{\bibfnamefont{W.}~\bibnamefont{Chen}},
  \bibinfo{author}{\bibfnamefont{T.}~\bibnamefont{Jamneala}},
  \bibinfo{author}{\bibfnamefont{M.~F.} \bibnamefont{Crommie}},
  \bibnamefont{and} \bibinfo{author}{\bibfnamefont{N.~S.}
  \bibnamefont{Wingreen}}, \bibinfo{journal}{Phys. Rev. B}
  \textbf{\bibinfo{volume}{64}}, \bibinfo{pages}{165412}
  (\bibinfo{year}{2001}).

\bibitem[{\citenamefont{Knorr et~al.}(2002)\citenamefont{Knorr, Schneider,
  Diekh\"oner, Wahl, and Kern}}]{knorr02}
\bibinfo{author}{\bibfnamefont{N.}~\bibnamefont{Knorr}},
  \bibinfo{author}{\bibfnamefont{M.~A.} \bibnamefont{Schneider}},
  \bibinfo{author}{\bibfnamefont{L.}~\bibnamefont{Diekh\"oner}},
  \bibinfo{author}{\bibfnamefont{P.}~\bibnamefont{Wahl}}, \bibnamefont{and}
  \bibinfo{author}{\bibfnamefont{K.}~\bibnamefont{Kern}},
  \bibinfo{journal}{Phys. Rev. Lett.} \textbf{\bibinfo{volume}{88}},
  \bibinfo{pages}{096804} (\bibinfo{year}{2002}).

\bibitem[{\citenamefont{Kouwenhoven and Glazman}(2001)}]{kouwenhoven01}
\bibinfo{author}{\bibfnamefont{L.}~\bibnamefont{Kouwenhoven}} \bibnamefont{and}
  \bibinfo{author}{\bibfnamefont{L.}~\bibnamefont{Glazman}},
  \bibinfo{journal}{Physics World} \textbf{\bibinfo{volume}{14}},
  \bibinfo{pages}{33} (\bibinfo{year}{2001}).

\bibitem[{\citenamefont{Henzl and Morgenstern}(2007)}]{henzl07}
\bibinfo{author}{\bibfnamefont{J.}~\bibnamefont{Henzl}} \bibnamefont{and}
  \bibinfo{author}{\bibfnamefont{K.}~\bibnamefont{Morgenstern}},
  \bibinfo{journal}{Phys. Rev. Lett.} \textbf{\bibinfo{volume}{98}},
  \bibinfo{pages}{266601} (\bibinfo{year}{2007}).

\bibitem[{\citenamefont{Franke et~al.}(2011)\citenamefont{Franke, Schulze, and
  Pascual}}]{franke11}
\bibinfo{author}{\bibfnamefont{K.}~\bibnamefont{Franke}},
  \bibinfo{author}{\bibfnamefont{G.}~\bibnamefont{Schulze}}, \bibnamefont{and}
  \bibinfo{author}{\bibfnamefont{J.}~\bibnamefont{Pascual}},
  \bibinfo{journal}{Science} \textbf{\bibinfo{volume}{332}},
  \bibinfo{pages}{940} (\bibinfo{year}{2011}).

\bibitem[{\citenamefont{Spinelli et~al.}(2015)\citenamefont{Spinelli, Gerrits,
  Toskovic, Bryant, Ternes, and Otte}}]{spinelli15}
\bibinfo{author}{\bibfnamefont{A.}~\bibnamefont{Spinelli}},
  \bibinfo{author}{\bibfnamefont{M.}~\bibnamefont{Gerrits}},
  \bibinfo{author}{\bibfnamefont{R.}~\bibnamefont{Toskovic}},
  \bibinfo{author}{\bibfnamefont{B.}~\bibnamefont{Bryant}},
  \bibinfo{author}{\bibfnamefont{M.}~\bibnamefont{Ternes}}, \bibnamefont{and}
  \bibinfo{author}{\bibfnamefont{A.}~\bibnamefont{Otte}},
  \bibinfo{journal}{Nature Commun.} \textbf{\bibinfo{volume}{4}},
  \bibinfo{pages}{10046} (\bibinfo{year}{2015}).

\bibitem[{\citenamefont{Mart{\'\i}nez-Velarte
  et~al.}(2017)\citenamefont{Mart{\'\i}nez-Velarte, Kretz, Moro-Lagares,
  Aguirre, Riedemann, Lograsso, Morellon, Ibarra, Garcia-Lekue, and
  Serrate}}]{martinez17}
\bibinfo{author}{\bibfnamefont{M.~C.} \bibnamefont{Mart{\'\i}nez-Velarte}},
  \bibinfo{author}{\bibfnamefont{B.}~\bibnamefont{Kretz}},
  \bibinfo{author}{\bibfnamefont{M.}~\bibnamefont{Moro-Lagares}},
  \bibinfo{author}{\bibfnamefont{M.~H.} \bibnamefont{Aguirre}},
  \bibinfo{author}{\bibfnamefont{T.~M.} \bibnamefont{Riedemann}},
  \bibinfo{author}{\bibfnamefont{T.~A.} \bibnamefont{Lograsso}},
  \bibinfo{author}{\bibfnamefont{L.}~\bibnamefont{Morellon}},
  \bibinfo{author}{\bibfnamefont{M.~R.} \bibnamefont{Ibarra}},
  \bibinfo{author}{\bibfnamefont{A.}~\bibnamefont{Garcia-Lekue}},
  \bibnamefont{and} \bibinfo{author}{\bibfnamefont{D.}~\bibnamefont{Serrate}},
  \bibinfo{journal}{Nano Letters}  (\bibinfo{year}{2017}).

\bibitem[{\citenamefont{Cornils et~al.}(2017)\citenamefont{Cornils, Kamlapure,
  Zhou, Pradhan, Khajetoorians, Fransson, Wiebe, and Wiesendanger}}]{cornils17}
\bibinfo{author}{\bibfnamefont{L.}~\bibnamefont{Cornils}},
  \bibinfo{author}{\bibfnamefont{A.}~\bibnamefont{Kamlapure}},
  \bibinfo{author}{\bibfnamefont{L.}~\bibnamefont{Zhou}},
  \bibinfo{author}{\bibfnamefont{S.}~\bibnamefont{Pradhan}},
  \bibinfo{author}{\bibfnamefont{A.}~\bibnamefont{Khajetoorians}},
  \bibinfo{author}{\bibfnamefont{J.}~\bibnamefont{Fransson}},
  \bibinfo{author}{\bibfnamefont{J.}~\bibnamefont{Wiebe}}, \bibnamefont{and}
  \bibinfo{author}{\bibfnamefont{R.}~\bibnamefont{Wiesendanger}},
  \bibinfo{journal}{Physical review letters} \textbf{\bibinfo{volume}{119}},
  \bibinfo{pages}{197002} (\bibinfo{year}{2017}).

\bibitem[{\citenamefont{Kondo}(1968)}]{kondo68}
\bibinfo{author}{\bibfnamefont{J.}~\bibnamefont{Kondo}},
  \bibinfo{journal}{Phys. Rev.} \textbf{\bibinfo{volume}{169}},
  \bibinfo{pages}{437} (\bibinfo{year}{1968}).

\bibitem[{\citenamefont{Hewson}(1997)}]{hewson97}
\bibinfo{author}{\bibfnamefont{A.~C.} \bibnamefont{Hewson}},
  \emph{\bibinfo{title}{The Kondo Problem to Heavy Fermions}}
  (\bibinfo{year}{1997}), ISBN \bibinfo{isbn}{9780521599474}.

\bibitem[{\citenamefont{Andres et~al.}(1975)\citenamefont{Andres, Graebner, and
  Ott}}]{andres75}
\bibinfo{author}{\bibfnamefont{K.}~\bibnamefont{Andres}},
  \bibinfo{author}{\bibfnamefont{J.~E.} \bibnamefont{Graebner}},
  \bibnamefont{and} \bibinfo{author}{\bibfnamefont{H.~R.} \bibnamefont{Ott}},
  \bibinfo{journal}{Phys. Rev. Lett.} \textbf{\bibinfo{volume}{35}},
  \bibinfo{pages}{1779} (\bibinfo{year}{1975}),
  \urlprefix\url{https://link.aps.org/doi/10.1103/PhysRevLett.35.1779}.

\bibitem[{\citenamefont{Aeppli and Fisk}(1992)}]{aeppli92}
\bibinfo{author}{\bibfnamefont{G.}~\bibnamefont{Aeppli}} \bibnamefont{and}
  \bibinfo{author}{\bibfnamefont{Z.}~\bibnamefont{Fisk}},
  \bibinfo{journal}{Comments Condens. Matter Phys.}
  \textbf{\bibinfo{volume}{16}}, \bibinfo{pages}{1192} (\bibinfo{year}{1992}).

\bibitem[{\citenamefont{Roch et~al.}(2008)\citenamefont{Roch, Florens,
  Bouchiat, Wernsdorfer, and Balestro}}]{roch08}
\bibinfo{author}{\bibfnamefont{N.}~\bibnamefont{Roch}},
  \bibinfo{author}{\bibfnamefont{S.}~\bibnamefont{Florens}},
  \bibinfo{author}{\bibfnamefont{V.}~\bibnamefont{Bouchiat}},
  \bibinfo{author}{\bibfnamefont{W.}~\bibnamefont{Wernsdorfer}},
  \bibnamefont{and} \bibinfo{author}{\bibfnamefont{F.}~\bibnamefont{Balestro}},
  \bibinfo{journal}{Nature} \textbf{\bibinfo{volume}{453}},
  \bibinfo{pages}{633} (\bibinfo{year}{2008}).

\bibitem[{\citenamefont{Parks et~al.}(2010)\citenamefont{Parks, Champagne,
  Costi, Shum, Pasupathy, Neuscamman, Flores-Torres, Cornaglia, Aligia,
  Balseiro et~al.}}]{parks10}
\bibinfo{author}{\bibfnamefont{J.~J.} \bibnamefont{Parks}},
  \bibinfo{author}{\bibfnamefont{A.~R.} \bibnamefont{Champagne}},
  \bibinfo{author}{\bibfnamefont{T.~A.} \bibnamefont{Costi}},
  \bibinfo{author}{\bibfnamefont{W.~W.} \bibnamefont{Shum}},
  \bibinfo{author}{\bibfnamefont{A.~N.} \bibnamefont{Pasupathy}},
  \bibinfo{author}{\bibfnamefont{E.}~\bibnamefont{Neuscamman}},
  \bibinfo{author}{\bibfnamefont{S.}~\bibnamefont{Flores-Torres}},
  \bibinfo{author}{\bibfnamefont{P.~S.} \bibnamefont{Cornaglia}},
  \bibinfo{author}{\bibfnamefont{A.~A.} \bibnamefont{Aligia}},
  \bibinfo{author}{\bibfnamefont{C.~A.} \bibnamefont{Balseiro}},
  \bibnamefont{et~al.}, \bibinfo{journal}{Science}
  \textbf{\bibinfo{volume}{328}}, \bibinfo{pages}{1370} (\bibinfo{year}{2010}),
  \urlprefix\url{http://www.sciencemag.org/content/328/5984/1370.abstract}.

\bibitem[{\citenamefont{Florens et~al.}(2011)\citenamefont{Florens, Freyn,
  Roch, Wernsdorfer, Balestro, Roura-Bas, and Aligia}}]{florens11}
\bibinfo{author}{\bibfnamefont{S.}~\bibnamefont{Florens}},
  \bibinfo{author}{\bibfnamefont{A.}~\bibnamefont{Freyn}},
  \bibinfo{author}{\bibfnamefont{N.}~\bibnamefont{Roch}},
  \bibinfo{author}{\bibfnamefont{W.}~\bibnamefont{Wernsdorfer}},
  \bibinfo{author}{\bibfnamefont{F.}~\bibnamefont{Balestro}},
  \bibinfo{author}{\bibfnamefont{P.}~\bibnamefont{Roura-Bas}},
  \bibnamefont{and} \bibinfo{author}{\bibfnamefont{A.}~\bibnamefont{Aligia}},
  \bibinfo{journal}{Journal of Physics: Condensed Matter}
  \textbf{\bibinfo{volume}{23}}, \bibinfo{pages}{243202}
  (\bibinfo{year}{2011}).

\bibitem[{\citenamefont{Vincent et~al.}(2012)\citenamefont{Vincent, Klyatskaya,
  Ruben, Wernsdorfer, and Balestro}}]{vincent12}
\bibinfo{author}{\bibfnamefont{R.}~\bibnamefont{Vincent}},
  \bibinfo{author}{\bibfnamefont{S.}~\bibnamefont{Klyatskaya}},
  \bibinfo{author}{\bibfnamefont{M.}~\bibnamefont{Ruben}},
  \bibinfo{author}{\bibfnamefont{W.}~\bibnamefont{Wernsdorfer}},
  \bibnamefont{and} \bibinfo{author}{\bibfnamefont{F.}~\bibnamefont{Balestro}},
  \bibinfo{journal}{Nature} \textbf{\bibinfo{volume}{488}},
  \bibinfo{pages}{357} (\bibinfo{year}{2012}).

\bibitem[{\citenamefont{Li et~al.}(1998)\citenamefont{Li, Schneider, Berndt,
  and Delley}}]{li98}
\bibinfo{author}{\bibfnamefont{J.}~\bibnamefont{Li}},
  \bibinfo{author}{\bibfnamefont{W.-D.} \bibnamefont{Schneider}},
  \bibinfo{author}{\bibfnamefont{R.}~\bibnamefont{Berndt}}, \bibnamefont{and}
  \bibinfo{author}{\bibfnamefont{B.}~\bibnamefont{Delley}},
  \bibinfo{journal}{Physical Review Letters} \textbf{\bibinfo{volume}{80}},
  \bibinfo{pages}{2893} (\bibinfo{year}{1998}).

\bibitem[{\citenamefont{Madhavan et~al.}(1998)\citenamefont{Madhavan, Chen,
  Jamneala, Crommie, and Wingreen}}]{madhavan98}
\bibinfo{author}{\bibfnamefont{V.}~\bibnamefont{Madhavan}},
  \bibinfo{author}{\bibfnamefont{W.}~\bibnamefont{Chen}},
  \bibinfo{author}{\bibfnamefont{T.}~\bibnamefont{Jamneala}},
  \bibinfo{author}{\bibfnamefont{M.~F.} \bibnamefont{Crommie}},
  \bibnamefont{and} \bibinfo{author}{\bibfnamefont{N.~S.}
  \bibnamefont{Wingreen}}, \bibinfo{journal}{Science}
  \textbf{\bibinfo{volume}{280}}, \bibinfo{pages}{567} (\bibinfo{year}{1998}),
  \urlprefix\url{http://www.sciencemag.org/content/280/5363/567.abstract}.

\bibitem[{\citenamefont{Manoharan et~al.}(2000)\citenamefont{Manoharan, Lutz,
  and Eigler}}]{manoharan00}
\bibinfo{author}{\bibfnamefont{H.~C.} \bibnamefont{Manoharan}},
  \bibinfo{author}{\bibfnamefont{C.~P.} \bibnamefont{Lutz}}, \bibnamefont{and}
  \bibinfo{author}{\bibfnamefont{D.~M.} \bibnamefont{Eigler}},
  \bibinfo{journal}{Nature} \textbf{\bibinfo{volume}{403}},
  \bibinfo{pages}{512} (\bibinfo{year}{2000}).

\bibitem[{\citenamefont{Limot et~al.}(2005)\citenamefont{Limot, Pehlke,
  Kr\"oger, and Berndt}}]{limot05}
\bibinfo{author}{\bibfnamefont{L.}~\bibnamefont{Limot}},
  \bibinfo{author}{\bibfnamefont{E.}~\bibnamefont{Pehlke}},
  \bibinfo{author}{\bibfnamefont{J.}~\bibnamefont{Kr\"oger}}, \bibnamefont{and}
  \bibinfo{author}{\bibfnamefont{R.}~\bibnamefont{Berndt}},
  \bibinfo{journal}{Phys. Rev. Lett.} \textbf{\bibinfo{volume}{94}},
  \bibinfo{pages}{036805} (\bibinfo{year}{2005}),
  \urlprefix\url{http://link.aps.org/doi/10.1103/PhysRevLett.94.036805}.

\bibitem[{\citenamefont{Serrate et~al.}(2014)\citenamefont{Serrate,
  Moro-Lagares, Piantek, Pascual, and Ibarra}}]{serrate14}
\bibinfo{author}{\bibfnamefont{D.}~\bibnamefont{Serrate}},
  \bibinfo{author}{\bibfnamefont{M.}~\bibnamefont{Moro-Lagares}},
  \bibinfo{author}{\bibfnamefont{M.}~\bibnamefont{Piantek}},
  \bibinfo{author}{\bibfnamefont{J.~I.} \bibnamefont{Pascual}},
  \bibnamefont{and} \bibinfo{author}{\bibfnamefont{M.~R.}
  \bibnamefont{Ibarra}}, \bibinfo{journal}{The Journal of Physical Chemistry C}
  \textbf{\bibinfo{volume}{118}}, \bibinfo{pages}{5827} (\bibinfo{year}{2014}).

\bibitem[{\citenamefont{Zhao et~al.}(2005)\citenamefont{Zhao, Li, Chen, Xiang,
  Wang, Pan, Wang, Xiao, Yang, Hou et~al.}}]{Zhao05}
\bibinfo{author}{\bibfnamefont{A.}~\bibnamefont{Zhao}},
  \bibinfo{author}{\bibfnamefont{Q.}~\bibnamefont{Li}},
  \bibinfo{author}{\bibfnamefont{L.}~\bibnamefont{Chen}},
  \bibinfo{author}{\bibfnamefont{H.}~\bibnamefont{Xiang}},
  \bibinfo{author}{\bibfnamefont{W.}~\bibnamefont{Wang}},
  \bibinfo{author}{\bibfnamefont{S.}~\bibnamefont{Pan}},
  \bibinfo{author}{\bibfnamefont{B.}~\bibnamefont{Wang}},
  \bibinfo{author}{\bibfnamefont{X.}~\bibnamefont{Xiao}},
  \bibinfo{author}{\bibfnamefont{J.}~\bibnamefont{Yang}},
  \bibinfo{author}{\bibfnamefont{J.~G.} \bibnamefont{Hou}},
  \bibnamefont{et~al.}, \bibinfo{journal}{Science}
  \textbf{\bibinfo{volume}{309}}, \bibinfo{pages}{1542} (\bibinfo{year}{2005}).

\bibitem[{\citenamefont{Komeda et~al.}(2011)\citenamefont{Komeda, Isshiki, Liu,
  Zhang, Lorente, Katoh, Breedlove, and Yamashita}}]{komeda11}
\bibinfo{author}{\bibfnamefont{T.}~\bibnamefont{Komeda}},
  \bibinfo{author}{\bibfnamefont{H.}~\bibnamefont{Isshiki}},
  \bibinfo{author}{\bibfnamefont{J.}~\bibnamefont{Liu}},
  \bibinfo{author}{\bibfnamefont{Y.-F.} \bibnamefont{Zhang}},
  \bibinfo{author}{\bibfnamefont{N.}~\bibnamefont{Lorente}},
  \bibinfo{author}{\bibfnamefont{K.}~\bibnamefont{Katoh}},
  \bibinfo{author}{\bibfnamefont{B.~K.} \bibnamefont{Breedlove}},
  \bibnamefont{and}
  \bibinfo{author}{\bibfnamefont{M.}~\bibnamefont{Yamashita}},
  \bibinfo{journal}{Nature communications} \textbf{\bibinfo{volume}{2}},
  \bibinfo{pages}{217} (\bibinfo{year}{2011}).

\bibitem[{\citenamefont{Minamitani et~al.}(2012)\citenamefont{Minamitani,
  Tsukahara, Matsunaka, Kim, Takagi, and Kawai}}]{minamitani12}
\bibinfo{author}{\bibfnamefont{E.}~\bibnamefont{Minamitani}},
  \bibinfo{author}{\bibfnamefont{N.}~\bibnamefont{Tsukahara}},
  \bibinfo{author}{\bibfnamefont{D.}~\bibnamefont{Matsunaka}},
  \bibinfo{author}{\bibfnamefont{Y.}~\bibnamefont{Kim}},
  \bibinfo{author}{\bibfnamefont{N.}~\bibnamefont{Takagi}}, \bibnamefont{and}
  \bibinfo{author}{\bibfnamefont{M.}~\bibnamefont{Kawai}},
  \bibinfo{journal}{Physical review letters} \textbf{\bibinfo{volume}{109}},
  \bibinfo{pages}{086602} (\bibinfo{year}{2012}).

\bibitem[{\citenamefont{Iancu et~al.}(2016)\citenamefont{Iancu, Schouteden, Li,
  and Van~Haesendonck}}]{iancu16}
\bibinfo{author}{\bibfnamefont{V.}~\bibnamefont{Iancu}},
  \bibinfo{author}{\bibfnamefont{K.}~\bibnamefont{Schouteden}},
  \bibinfo{author}{\bibfnamefont{Z.}~\bibnamefont{Li}}, \bibnamefont{and}
  \bibinfo{author}{\bibfnamefont{C.}~\bibnamefont{Van~Haesendonck}},
  \bibinfo{journal}{Chemical Communications} \textbf{\bibinfo{volume}{52}},
  \bibinfo{pages}{11359} (\bibinfo{year}{2016}).

\bibitem[{\citenamefont{Ormaza et~al.}(2017)\citenamefont{Ormaza, Abufager,
  Verlhac, Bachellier, Bocquet, Lorente, and Limot}}]{ormaza17}
\bibinfo{author}{\bibfnamefont{M.}~\bibnamefont{Ormaza}},
  \bibinfo{author}{\bibfnamefont{P.}~\bibnamefont{Abufager}},
  \bibinfo{author}{\bibfnamefont{B.}~\bibnamefont{Verlhac}},
  \bibinfo{author}{\bibfnamefont{N.}~\bibnamefont{Bachellier}},
  \bibinfo{author}{\bibfnamefont{M.-L.} \bibnamefont{Bocquet}},
  \bibinfo{author}{\bibfnamefont{N.}~\bibnamefont{Lorente}}, \bibnamefont{and}
  \bibinfo{author}{\bibfnamefont{L.}~\bibnamefont{Limot}},
  \bibinfo{journal}{Nature communications} \textbf{\bibinfo{volume}{8}},
  \bibinfo{pages}{1974} (\bibinfo{year}{2017}).

\bibitem[{\citenamefont{Fern\'andez et~al.}(2016)\citenamefont{Fern\'andez,
  Moro-Lagares, Serrate, and Aligia}}]{fernandez16}
\bibinfo{author}{\bibfnamefont{J.}~\bibnamefont{Fern\'andez}},
  \bibinfo{author}{\bibfnamefont{M.}~\bibnamefont{Moro-Lagares}},
  \bibinfo{author}{\bibfnamefont{D.}~\bibnamefont{Serrate}}, \bibnamefont{and}
  \bibinfo{author}{\bibfnamefont{A.~A.} \bibnamefont{Aligia}},
  \bibinfo{journal}{Phys. Rev. B} \textbf{\bibinfo{volume}{94}},
  \bibinfo{pages}{075408} (\bibinfo{year}{2016}).

\bibitem[{\citenamefont{\'Ujs\'aghy et~al.}(2000)\citenamefont{\'Ujs\'aghy,
  Kroha, Szunyogh, and Zawadowski}}]{ujsaghy00}
\bibinfo{author}{\bibfnamefont{O.}~\bibnamefont{\'Ujs\'aghy}},
  \bibinfo{author}{\bibfnamefont{J.}~\bibnamefont{Kroha}},
  \bibinfo{author}{\bibfnamefont{L.}~\bibnamefont{Szunyogh}}, \bibnamefont{and}
  \bibinfo{author}{\bibfnamefont{A.}~\bibnamefont{Zawadowski}},
  \bibinfo{journal}{Phys. Rev. Lett.} \textbf{\bibinfo{volume}{85}},
  \bibinfo{pages}{2557} (\bibinfo{year}{2000}),
  \urlprefix\url{http://link.aps.org/doi/10.1103/PhysRevLett.85.2557}.

\bibitem[{\citenamefont{Plihal and Gadzuk}(2001)}]{plihal01}
\bibinfo{author}{\bibfnamefont{M.}~\bibnamefont{Plihal}} \bibnamefont{and}
  \bibinfo{author}{\bibfnamefont{J.~W.} \bibnamefont{Gadzuk}},
  \bibinfo{journal}{Phys. Rev. B} \textbf{\bibinfo{volume}{63}},
  \bibinfo{pages}{085404} (\bibinfo{year}{2001}),
  \urlprefix\url{http://link.aps.org/doi/10.1103/PhysRevB.63.085404}.

\bibitem[{\citenamefont{Schneider et~al.}(2005)\citenamefont{Schneider, Wahl,
  Diekh\"{o}ner, Vitali, Wittich, and Kern}}]{schneider05}
\bibinfo{author}{\bibfnamefont{M.~A.} \bibnamefont{Schneider}},
  \bibinfo{author}{\bibfnamefont{P.}~\bibnamefont{Wahl}},
  \bibinfo{author}{\bibfnamefont{L.}~\bibnamefont{Diekh\"{o}ner}},
  \bibinfo{author}{\bibfnamefont{L.}~\bibnamefont{Vitali}},
  \bibinfo{author}{\bibfnamefont{G.}~\bibnamefont{Wittich}}, \bibnamefont{and}
  \bibinfo{author}{\bibfnamefont{K.}~\bibnamefont{Kern}},
  \bibinfo{journal}{Japanese Journal of Applied Physics}
  \textbf{\bibinfo{volume}{44}}, \bibinfo{pages}{5328} (\bibinfo{year}{2005}),
  \urlprefix\url{http://jjap.jsap.jp/link?JJAP/44/5328/}.

\bibitem[{\citenamefont{Merino and Gunnarsson}(2004)}]{merino04}
\bibinfo{author}{\bibfnamefont{J.}~\bibnamefont{Merino}} \bibnamefont{and}
  \bibinfo{author}{\bibfnamefont{O.}~\bibnamefont{Gunnarsson}},
  \bibinfo{journal}{Phys. Rev. Lett.} \textbf{\bibinfo{volume}{93}},
  \bibinfo{pages}{156601} (\bibinfo{year}{2004}),
  \urlprefix\url{http://link.aps.org/doi/10.1103/PhysRevLett.93.156601}.

\bibitem[{\citenamefont{Aligia and Lobos}(2005)}]{aligia05}
\bibinfo{author}{\bibfnamefont{A.~A.} \bibnamefont{Aligia}} \bibnamefont{and}
  \bibinfo{author}{\bibfnamefont{A.~M.} \bibnamefont{Lobos}},
  \bibinfo{journal}{Journal of Physics: Condensed Matter}
  \textbf{\bibinfo{volume}{17}}, \bibinfo{pages}{S1095} (\bibinfo{year}{2005}),
  \urlprefix\url{http://stacks.iop.org/0953-8984/17/i=13/a=005}.

\bibitem[{\citenamefont{Li et~al.}(2009)\citenamefont{Li, Yamazaki, Eguchi,
  Kim, Kahng, Jia, Xue, and Hasegawa}}]{li09}
\bibinfo{author}{\bibfnamefont{Q.}~\bibnamefont{Li}},
  \bibinfo{author}{\bibfnamefont{S.}~\bibnamefont{Yamazaki}},
  \bibinfo{author}{\bibfnamefont{T.}~\bibnamefont{Eguchi}},
  \bibinfo{author}{\bibfnamefont{H.}~\bibnamefont{Kim}},
  \bibinfo{author}{\bibfnamefont{S.-J.} \bibnamefont{Kahng}},
  \bibinfo{author}{\bibfnamefont{J.~F.} \bibnamefont{Jia}},
  \bibinfo{author}{\bibfnamefont{Q.~K.} \bibnamefont{Xue}}, \bibnamefont{and}
  \bibinfo{author}{\bibfnamefont{Y.}~\bibnamefont{Hasegawa}},
  \bibinfo{journal}{Phys. Rev. B} \textbf{\bibinfo{volume}{80}},
  \bibinfo{pages}{115431} (\bibinfo{year}{2009}).

\bibitem[{\citenamefont{Li et~al.}(2018)\citenamefont{Li, Zheng, Wang, Miao,
  Cao, Sun, Wu, Wu, Li, Wang et~al.}}]{li18}
\bibinfo{author}{\bibfnamefont{Q.~L.} \bibnamefont{Li}},
  \bibinfo{author}{\bibfnamefont{C.}~\bibnamefont{Zheng}},
  \bibinfo{author}{\bibfnamefont{R.}~\bibnamefont{Wang}},
  \bibinfo{author}{\bibfnamefont{B.~F.} \bibnamefont{Miao}},
  \bibinfo{author}{\bibfnamefont{R.~X.} \bibnamefont{Cao}},
  \bibinfo{author}{\bibfnamefont{L.}~\bibnamefont{Sun}},
  \bibinfo{author}{\bibfnamefont{D.}~\bibnamefont{Wu}},
  \bibinfo{author}{\bibfnamefont{Y.~Z.} \bibnamefont{Wu}},
  \bibinfo{author}{\bibfnamefont{S.~C.} \bibnamefont{Li}},
  \bibinfo{author}{\bibfnamefont{B.~G.} \bibnamefont{Wang}},
  \bibnamefont{et~al.}, \bibinfo{journal}{Phys. Rev. B}
  \textbf{\bibinfo{volume}{97}}, \bibinfo{pages}{035417}
  (\bibinfo{year}{2018}),
  \urlprefix\url{https://link.aps.org/doi/10.1103/PhysRevB.97.035417}.

\bibitem[{\citenamefont{Lin et~al.}(2005)\citenamefont{Lin, Castro~Neto, and
  Jones}}]{lin05}
\bibinfo{author}{\bibfnamefont{C.-Y.} \bibnamefont{Lin}},
  \bibinfo{author}{\bibfnamefont{A.~H.} \bibnamefont{Castro~Neto}},
  \bibnamefont{and} \bibinfo{author}{\bibfnamefont{B.~A.} \bibnamefont{Jones}},
  \bibinfo{journal}{Phys. Rev. B} \textbf{\bibinfo{volume}{71}},
  \bibinfo{pages}{035417} (\bibinfo{year}{2005}),
  \urlprefix\url{https://link.aps.org/doi/10.1103/PhysRevB.71.035417}.

\bibitem[{\citenamefont{Ternes et~al.}(2008)\citenamefont{Ternes, Lutz,
  Hirjibehedin, Giessibl, and Heinrich}}]{ternes08}
\bibinfo{author}{\bibfnamefont{M.}~\bibnamefont{Ternes}},
  \bibinfo{author}{\bibfnamefont{C.~P.} \bibnamefont{Lutz}},
  \bibinfo{author}{\bibfnamefont{C.~F.} \bibnamefont{Hirjibehedin}},
  \bibinfo{author}{\bibfnamefont{F.~J.} \bibnamefont{Giessibl}},
  \bibnamefont{and} \bibinfo{author}{\bibfnamefont{A.~J.}
  \bibnamefont{Heinrich}}, \bibinfo{journal}{Science}
  \textbf{\bibinfo{volume}{319}}, \bibinfo{pages}{1066} (\bibinfo{year}{2008}).

\bibitem[{\citenamefont{Aligia}(2013)}]{aligia13}
\bibinfo{author}{\bibfnamefont{A.~A.} \bibnamefont{Aligia}},
  \bibinfo{journal}{Phys. Rev. B} \textbf{\bibinfo{volume}{88}},
  \bibinfo{pages}{075128} (\bibinfo{year}{2013}).

\bibitem[{not({\natexlab{a}})}]{note3}
\bibinfo{note}{In our own experiments as well as those of Ref.
  \onlinecite{li18} the dip in the differential conductance has a considerable
  shift to the right of the Fermi level. This is consistent with a Kondo effect
  with total d occupancy $n_d$ near 1 for SU(4) symmetry but is not consistent
  with either $n_d \approx 1$ and SU(2) symmetry or $n_d \approx 2$ for the
  two-channel model. In these two cases, the Kondo peak is practically at the
  Fermi level in the Kondo limit [\onlinecite{tosi12}]. In the supplemental
  material of Ref. \onlinecite{li18} the Friedel sume rule
  [\onlinecite{tosi12}] is inverted to estimate $0.4 < n_d <0.66$ based on the
  simplest one-channel SU(2) Anderson model. This would indicate that the
  system is in the intermediate valence regime, instead of the Kondo one. The
  same analysis for the SU(4) case gives $0.8 < n_d < 1.3$ which is fully
  consistent with our SU(4) model in the Kondo regime.}

\bibitem[{\citenamefont{Barral et~al.}(2017)\citenamefont{Barral, Di~Napoli,
  Blesio, Roura-Bas, Camjayi, Manuel, and Aligia}}]{barral17}
\bibinfo{author}{\bibfnamefont{M.}~\bibnamefont{Barral}},
  \bibinfo{author}{\bibfnamefont{S.}~\bibnamefont{Di~Napoli}},
  \bibinfo{author}{\bibfnamefont{G.}~\bibnamefont{Blesio}},
  \bibinfo{author}{\bibfnamefont{P.}~\bibnamefont{Roura-Bas}},
  \bibinfo{author}{\bibfnamefont{A.}~\bibnamefont{Camjayi}},
  \bibinfo{author}{\bibfnamefont{L.}~\bibnamefont{Manuel}}, \bibnamefont{and}
  \bibinfo{author}{\bibfnamefont{A.}~\bibnamefont{Aligia}},
  \bibinfo{journal}{The Journal of Chemical Physics}
  \textbf{\bibinfo{volume}{146}}, \bibinfo{pages}{092315}
  (\bibinfo{year}{2017}).

\bibitem[{not({\natexlab{b}})}]{note2}
\bibinfo{note}{Calculations in a similar model, indicate that the main effect
  of a finite $U$ is a shift lo lower energies of the Kondo dip and an increase
  of its width that can be absorbed renormalizing the (unknown) magnitude of
  both surface and bulk hybriditations by the same factor
  [\citenum{fernandez15}]}.

\bibitem[{\citenamefont{Li et~al.}(1997)\citenamefont{Li, Schneider, and
  Berndt}}]{li97}
\bibinfo{author}{\bibfnamefont{J.}~\bibnamefont{Li}},
  \bibinfo{author}{\bibfnamefont{W.-D.} \bibnamefont{Schneider}},
  \bibnamefont{and} \bibinfo{author}{\bibfnamefont{R.}~\bibnamefont{Berndt}},
  \bibinfo{journal}{Phys. Rev. B} \textbf{\bibinfo{volume}{56}},
  \bibinfo{pages}{7656} (\bibinfo{year}{1997}),
  \urlprefix\url{https://link.aps.org/doi/10.1103/PhysRevB.56.7656}.

\bibitem[{\citenamefont{Bickers}(1987)}]{bickers87}
\bibinfo{author}{\bibfnamefont{N.}~\bibnamefont{Bickers}},
  \bibinfo{journal}{Reviews of modern physics} \textbf{\bibinfo{volume}{59}},
  \bibinfo{pages}{845} (\bibinfo{year}{1987}).

\bibitem[{\citenamefont{Anderson}(1970)}]{anderson70}
\bibinfo{author}{\bibfnamefont{P.}~\bibnamefont{Anderson}},
  \bibinfo{journal}{Journal of Physics C: Solid State Physics}
  \textbf{\bibinfo{volume}{3}}, \bibinfo{pages}{2436} (\bibinfo{year}{1970}).

\bibitem[{\citenamefont{\ifmmode~\check{Z}\else
  \v{Z}\fi{}itko}(2011{\natexlab{a}})}]{zitko11rg}
\bibinfo{author}{\bibfnamefont{R.}~\bibnamefont{\ifmmode~\check{Z}\else
  \v{Z}\fi{}itko}}, \bibinfo{journal}{Phys. Rev. B}
  \textbf{\bibinfo{volume}{84}}, \bibinfo{pages}{085142}
  (\bibinfo{year}{2011}{\natexlab{a}}),
  \urlprefix\url{https://link.aps.org/doi/10.1103/PhysRevB.84.085142}.

\bibitem[{\citenamefont{Vaugier et~al.}(2007)\citenamefont{Vaugier, Aligia, and
  Lobos}}]{vaugier07}
\bibinfo{author}{\bibfnamefont{L.}~\bibnamefont{Vaugier}},
  \bibinfo{author}{\bibfnamefont{A.}~\bibnamefont{Aligia}}, \bibnamefont{and}
  \bibinfo{author}{\bibfnamefont{A.}~\bibnamefont{Lobos}},
  \bibinfo{journal}{Physical Review B} \textbf{\bibinfo{volume}{76}},
  \bibinfo{pages}{165112} (\bibinfo{year}{2007}).

\bibitem[{\citenamefont{\ifmmode~\check{Z}\else
  \v{Z}\fi{}itko}(2011{\natexlab{b}})}]{zitko11}
\bibinfo{author}{\bibfnamefont{R.}~\bibnamefont{\ifmmode~\check{Z}\else
  \v{Z}\fi{}itko}}, \bibinfo{journal}{Phys. Rev. B}
  \textbf{\bibinfo{volume}{84}}, \bibinfo{pages}{195116}
  (\bibinfo{year}{2011}{\natexlab{b}}),
  \urlprefix\url{http://link.aps.org/doi/10.1103/PhysRevB.84.195116}.

\bibitem[{\citenamefont{Aligia et~al.}(2015)\citenamefont{Aligia, Roura-Bas,
  and Florens}}]{aligia15}
\bibinfo{author}{\bibfnamefont{A.}~\bibnamefont{Aligia}},
  \bibinfo{author}{\bibfnamefont{P.}~\bibnamefont{Roura-Bas}},
  \bibnamefont{and} \bibinfo{author}{\bibfnamefont{S.}~\bibnamefont{Florens}},
  \bibinfo{journal}{Physical Review B} \textbf{\bibinfo{volume}{92}},
  \bibinfo{pages}{035404} (\bibinfo{year}{2015}).

\bibitem[{\citenamefont{Fern\'andez et~al.}(2018)\citenamefont{Fern\'andez,
  Lisandrini, Roura-Bas, Gazza, and Aligia}}]{fernandez18}
\bibinfo{author}{\bibfnamefont{J.}~\bibnamefont{Fern\'andez}},
  \bibinfo{author}{\bibfnamefont{F.}~\bibnamefont{Lisandrini}},
  \bibinfo{author}{\bibfnamefont{P.}~\bibnamefont{Roura-Bas}},
  \bibinfo{author}{\bibfnamefont{C.}~\bibnamefont{Gazza}}, \bibnamefont{and}
  \bibinfo{author}{\bibfnamefont{A.~A.} \bibnamefont{Aligia}},
  \bibinfo{journal}{Phys. Rev. B} \textbf{\bibinfo{volume}{97}},
  \bibinfo{pages}{045144} (\bibinfo{year}{2018}),
  \urlprefix\url{https://link.aps.org/doi/10.1103/PhysRevB.97.045144}.

\bibitem[{\citenamefont{Pruschke and Grewe}(1989)}]{pruschke89}
\bibinfo{author}{\bibfnamefont{T.}~\bibnamefont{Pruschke}} \bibnamefont{and}
  \bibinfo{author}{\bibfnamefont{N.}~\bibnamefont{Grewe}},
  \bibinfo{journal}{Zeitschrift f{\"u}r Physik B Condensed Matter}
  \textbf{\bibinfo{volume}{74}}, \bibinfo{pages}{439} (\bibinfo{year}{1989}).

\bibitem[{\citenamefont{Logan et~al.}(1998)\citenamefont{Logan, Eastwood, and
  Tusch}}]{logan98}
\bibinfo{author}{\bibfnamefont{D.~E.} \bibnamefont{Logan}},
  \bibinfo{author}{\bibfnamefont{M.~P.} \bibnamefont{Eastwood}},
  \bibnamefont{and} \bibinfo{author}{\bibfnamefont{M.~A.} \bibnamefont{Tusch}},
  \bibinfo{journal}{Journal of Physics: Condensed Matter}
  \textbf{\bibinfo{volume}{10}}, \bibinfo{pages}{2673} (\bibinfo{year}{1998}).

\bibitem[{\citenamefont{K{\"o}nemann et~al.}(2006)\citenamefont{K{\"o}nemann,
  Kubala, K{\"o}nig, and Haug}}]{konemann06}
\bibinfo{author}{\bibfnamefont{J.}~\bibnamefont{K{\"o}nemann}},
  \bibinfo{author}{\bibfnamefont{B.}~\bibnamefont{Kubala}},
  \bibinfo{author}{\bibfnamefont{J.}~\bibnamefont{K{\"o}nig}},
  \bibnamefont{and} \bibinfo{author}{\bibfnamefont{R.~J.} \bibnamefont{Haug}},
  \bibinfo{journal}{Physical Review B} \textbf{\bibinfo{volume}{73}},
  \bibinfo{pages}{033313} (\bibinfo{year}{2006}).

\bibitem[{\citenamefont{Kroha et~al.}(1998)}]{kroha98}
\bibinfo{author}{\bibfnamefont{J.}~\bibnamefont{Kroha}} \bibnamefont{et~al.},
  \bibinfo{journal}{Acta Phys. Pol. B} \textbf{\bibinfo{volume}{29}},
  \bibinfo{pages}{3781} (\bibinfo{year}{1998}).

\bibitem[{\citenamefont{Fern\'andez et~al.}(2017)\citenamefont{Fern\'andez,
  Aligia, Roura-Bas, and Andrade}}]{fernandez17}
\bibinfo{author}{\bibfnamefont{J.}~\bibnamefont{Fern\'andez}},
  \bibinfo{author}{\bibfnamefont{A.~A.} \bibnamefont{Aligia}},
  \bibinfo{author}{\bibfnamefont{P.}~\bibnamefont{Roura-Bas}},
  \bibnamefont{and} \bibinfo{author}{\bibfnamefont{J.~A.}
  \bibnamefont{Andrade}}, \bibinfo{journal}{Phys. Rev. B}
  \textbf{\bibinfo{volume}{95}}, \bibinfo{pages}{045143}
  (\bibinfo{year}{2017}).

\bibitem[{\citenamefont{Wingreen and Meir}(1994)}]{wingreen94}
\bibinfo{author}{\bibfnamefont{N.~S.} \bibnamefont{Wingreen}} \bibnamefont{and}
  \bibinfo{author}{\bibfnamefont{Y.}~\bibnamefont{Meir}},
  \bibinfo{journal}{Physical review B} \textbf{\bibinfo{volume}{49}},
  \bibinfo{pages}{11040} (\bibinfo{year}{1994}).

\bibitem[{\citenamefont{Reinert et~al.}(2001)\citenamefont{Reinert, Ehm,
  Schmidt, Nicolay, H{\"u}fner, Kroha, Trovarelli, and Geibel}}]{reinert01}
\bibinfo{author}{\bibfnamefont{F.}~\bibnamefont{Reinert}},
  \bibinfo{author}{\bibfnamefont{D.}~\bibnamefont{Ehm}},
  \bibinfo{author}{\bibfnamefont{S.}~\bibnamefont{Schmidt}},
  \bibinfo{author}{\bibfnamefont{G.}~\bibnamefont{Nicolay}},
  \bibinfo{author}{\bibfnamefont{S.}~\bibnamefont{H{\"u}fner}},
  \bibinfo{author}{\bibfnamefont{J.}~\bibnamefont{Kroha}},
  \bibinfo{author}{\bibfnamefont{O.}~\bibnamefont{Trovarelli}},
  \bibnamefont{and} \bibinfo{author}{\bibfnamefont{C.}~\bibnamefont{Geibel}},
  \bibinfo{journal}{Physical review letters} \textbf{\bibinfo{volume}{87}},
  \bibinfo{pages}{106401} (\bibinfo{year}{2001}).

\bibitem[{\citenamefont{Ehm et~al.}(2007)\citenamefont{Ehm, H{\"u}fner,
  Reinert, Kroha, W{\"o}lfle, Stockert, Geibel, and L{\"o}hneysen}}]{ehm07}
\bibinfo{author}{\bibfnamefont{D.}~\bibnamefont{Ehm}},
  \bibinfo{author}{\bibfnamefont{S.}~\bibnamefont{H{\"u}fner}},
  \bibinfo{author}{\bibfnamefont{F.}~\bibnamefont{Reinert}},
  \bibinfo{author}{\bibfnamefont{J.}~\bibnamefont{Kroha}},
  \bibinfo{author}{\bibfnamefont{P.}~\bibnamefont{W{\"o}lfle}},
  \bibinfo{author}{\bibfnamefont{O.}~\bibnamefont{Stockert}},
  \bibinfo{author}{\bibfnamefont{C.}~\bibnamefont{Geibel}}, \bibnamefont{and}
  \bibinfo{author}{\bibfnamefont{H.~v.} \bibnamefont{L{\"o}hneysen}},
  \bibinfo{journal}{Physical Review B} \textbf{\bibinfo{volume}{76}},
  \bibinfo{pages}{045117} (\bibinfo{year}{2007}).

\bibitem[{\citenamefont{Tosi et~al.}(2015)\citenamefont{Tosi, Roura-Bas, and
  Aligia}}]{tosi15}
\bibinfo{author}{\bibfnamefont{L.}~\bibnamefont{Tosi}},
  \bibinfo{author}{\bibfnamefont{P.}~\bibnamefont{Roura-Bas}},
  \bibnamefont{and} \bibinfo{author}{\bibfnamefont{A.}~\bibnamefont{Aligia}},
  \bibinfo{journal}{Journal of Physics: Condensed Matter}
  \textbf{\bibinfo{volume}{27}}, \bibinfo{pages}{335601}
  (\bibinfo{year}{2015}).

\bibitem[{\citenamefont{Di~Napoli et~al.}(2014)\citenamefont{Di~Napoli,
  Roura-Bas, Weichselbaum, and Aligia}}]{dinapoli14}
\bibinfo{author}{\bibfnamefont{S.}~\bibnamefont{Di~Napoli}},
  \bibinfo{author}{\bibfnamefont{P.}~\bibnamefont{Roura-Bas}},
  \bibinfo{author}{\bibfnamefont{A.}~\bibnamefont{Weichselbaum}},
  \bibnamefont{and} \bibinfo{author}{\bibfnamefont{A.}~\bibnamefont{Aligia}},
  \bibinfo{journal}{Physical Review B} \textbf{\bibinfo{volume}{90}},
  \bibinfo{pages}{125149} (\bibinfo{year}{2014}).

\bibitem[{\citenamefont{Bas and Aligia}(2009)}]{rourabas09transp}
\bibinfo{author}{\bibfnamefont{P.~R.} \bibnamefont{Bas}} \bibnamefont{and}
  \bibinfo{author}{\bibfnamefont{A.}~\bibnamefont{Aligia}},
  \bibinfo{journal}{Physical Review B} \textbf{\bibinfo{volume}{80}},
  \bibinfo{pages}{035308} (\bibinfo{year}{2009}).

\bibitem[{\citenamefont{Roura-Bas and Aligia}(2009)}]{rourabas09dyn}
\bibinfo{author}{\bibfnamefont{P.}~\bibnamefont{Roura-Bas}} \bibnamefont{and}
  \bibinfo{author}{\bibfnamefont{A.~A.} \bibnamefont{Aligia}},
  \bibinfo{journal}{Journal of Physics: Condensed Matter}
  \textbf{\bibinfo{volume}{22}}, \bibinfo{pages}{025602}
  (\bibinfo{year}{2009}).

\bibitem[{\citenamefont{Haule et~al.}(2001)\citenamefont{Haule, Kirchner,
  Kroha, and W{\"o}lfle}}]{haule01}
\bibinfo{author}{\bibfnamefont{K.}~\bibnamefont{Haule}},
  \bibinfo{author}{\bibfnamefont{S.}~\bibnamefont{Kirchner}},
  \bibinfo{author}{\bibfnamefont{J.}~\bibnamefont{Kroha}}, \bibnamefont{and}
  \bibinfo{author}{\bibfnamefont{P.}~\bibnamefont{W{\"o}lfle}},
  \bibinfo{journal}{Physical Review B} \textbf{\bibinfo{volume}{64}},
  \bibinfo{pages}{155111} (\bibinfo{year}{2001}).

\bibitem[{\citenamefont{Tosi et~al.}(2011)\citenamefont{Tosi, Roura-Bas, Llois,
  and Manuel}}]{tosi11}
\bibinfo{author}{\bibfnamefont{L.}~\bibnamefont{Tosi}},
  \bibinfo{author}{\bibfnamefont{P.}~\bibnamefont{Roura-Bas}},
  \bibinfo{author}{\bibfnamefont{A.~M.} \bibnamefont{Llois}}, \bibnamefont{and}
  \bibinfo{author}{\bibfnamefont{L.~O.} \bibnamefont{Manuel}},
  \bibinfo{journal}{Physical Review B} \textbf{\bibinfo{volume}{83}},
  \bibinfo{pages}{073301} (\bibinfo{year}{2011}).

\bibitem[{\citenamefont{Schneider et~al.}(2002)\citenamefont{Schneider, Vitali,
  Knorr, and Kern}}]{schneider02}
\bibinfo{author}{\bibfnamefont{M.~A.} \bibnamefont{Schneider}},
  \bibinfo{author}{\bibfnamefont{L.}~\bibnamefont{Vitali}},
  \bibinfo{author}{\bibfnamefont{N.}~\bibnamefont{Knorr}}, \bibnamefont{and}
  \bibinfo{author}{\bibfnamefont{K.}~\bibnamefont{Kern}},
  \bibinfo{journal}{Phys. Rev. B} \textbf{\bibinfo{volume}{65}},
  \bibinfo{pages}{121406} (\bibinfo{year}{2002}),
  \urlprefix\url{http://link.aps.org/doi/10.1103/PhysRevB.65.121406}.

\bibitem[{\citenamefont{Nagaoka et~al.}(2002)\citenamefont{Nagaoka, Jamneala,
  Grobis, and Crommie}}]{nagaoka02}
\bibinfo{author}{\bibfnamefont{K.}~\bibnamefont{Nagaoka}},
  \bibinfo{author}{\bibfnamefont{T.}~\bibnamefont{Jamneala}},
  \bibinfo{author}{\bibfnamefont{M.}~\bibnamefont{Grobis}}, \bibnamefont{and}
  \bibinfo{author}{\bibfnamefont{M.~F.} \bibnamefont{Crommie}},
  \bibinfo{journal}{Phys. Rev. Lett.} \textbf{\bibinfo{volume}{88}},
  \bibinfo{pages}{077205} (\bibinfo{year}{2002}),
  \urlprefix\url{http://link.aps.org/doi/10.1103/PhysRevLett.88.077205}.

\bibitem[{\citenamefont{Moro-Lagares}(2017)}]{phd:moro}
\bibinfo{author}{\bibfnamefont{M.}~\bibnamefont{Moro-Lagares}},
  \emph{\bibinfo{title}{Engineering Spin Structures at the Atomic Scale}}
  (\bibinfo{publisher}{Prensas de la Universidad de Zaragoza},
  \bibinfo{year}{2017}), ISBN \bibinfo{isbn}{978-84-16935-83-3}.

\bibitem[{\citenamefont{Tosi et~al.}(2012)\citenamefont{Tosi, Roura-Bas, Llois,
  and Aligia}}]{tosi12}
\bibinfo{author}{\bibfnamefont{L.}~\bibnamefont{Tosi}},
  \bibinfo{author}{\bibfnamefont{P.}~\bibnamefont{Roura-Bas}},
  \bibinfo{author}{\bibfnamefont{A.}~\bibnamefont{Llois}}, \bibnamefont{and}
  \bibinfo{author}{\bibfnamefont{A.}~\bibnamefont{Aligia}},
  \bibinfo{journal}{Physica B: Condensed Matter}
  \textbf{\bibinfo{volume}{407}}, \bibinfo{pages}{3263} (\bibinfo{year}{2012}).

\bibitem[{\citenamefont{Hewson}(1993)}]{hewson93}
\bibinfo{author}{\bibfnamefont{A.~C.} \bibnamefont{Hewson}},
  \emph{\bibinfo{title}{The Kondo Problem to Heavy Fermions}}
  (\bibinfo{year}{1993}).

\bibitem[{\citenamefont{Yang et~al.}(2008)\citenamefont{Yang, Fisk, Lee,
  Thompson, and Pines}}]{yang08}
\bibinfo{author}{\bibfnamefont{Y.-F.} \bibnamefont{Yang}},
  \bibinfo{author}{\bibfnamefont{Z.}~\bibnamefont{Fisk}},
  \bibinfo{author}{\bibfnamefont{H.-O.} \bibnamefont{Lee}},
  \bibinfo{author}{\bibfnamefont{J.}~\bibnamefont{Thompson}}, \bibnamefont{and}
  \bibinfo{author}{\bibfnamefont{D.}~\bibnamefont{Pines}},
  \bibinfo{journal}{Nature} \textbf{\bibinfo{volume}{454}},
  \bibinfo{pages}{611} (\bibinfo{year}{2008}).

\bibitem[{\citenamefont{Fernández et~al.}(2015)\citenamefont{Fernández,
  Aligia, and Lobos}}]{fernandez15}
\bibinfo{author}{\bibfnamefont{J.}~\bibnamefont{Fernández}},
  \bibinfo{author}{\bibfnamefont{A.~A.} \bibnamefont{Aligia}},
  \bibnamefont{and} \bibinfo{author}{\bibfnamefont{A.~M.} \bibnamefont{Lobos}},
  \bibinfo{journal}{EPL (Europhysics Letters)} \textbf{\bibinfo{volume}{109}},
  \bibinfo{pages}{37011} (\bibinfo{year}{2015}).

\end{thebibliography}

\end{document}